\documentclass[showpacs,onecolumn,preprintnumbers,amsmath,amsfonts,amssymb,floatfix,aps,superscriptaddress]{revtex4}

\usepackage{graphicx}
\usepackage{epsfig}
\usepackage{mathtools}
\usepackage[hang,nooneline]{subfigure}
\usepackage[normalem]{ulem}
\usepackage{color}
\usepackage{hyperref}

\usepackage{times}

\newcounter{fig}

\DeclareMathOperator{\arccosh}{arccosh}
\DeclareMathOperator{\arcsinh}{arcsinh}

\begin{document}

\title{Kuznetsov-Ma breather-like solutions in the Salerno model}

\author{J. Sullivan}
\email[Email: ]{jksullivan@umass.edu}
\affiliation{Department of Mathematics and
Statistics, University of Massachusetts
Amherst, Amherst, MA 01003-4515, USA}

\author{E.~G. Charalampidis}
\email[Email: ]{echarala@calpoly.edu}
\affiliation{Mathematics Department,
California Polytechnic State University,
San Luis Obispo, CA 93407-0403, USA}

\author{J.~Cuevas-Maraver}
\email[Email: ]{jcuevas@us.es}
\affiliation{Grupo de F\'{i}sica No Lineal, Departamento de F\'{i}sica Aplicada I,
Universidad de Sevilla. Escuela Polit\'{e}cnica Superior, C/ Virgen de \'{A}frica,
7, 41011-Sevilla, Spain\\
Instituto de Matem\'{a}ticas de la Universidad de Sevilla (IMUS).
Edificio Celestino Mutis. Avda. Reina Mercedes s/n, 41012-Sevilla, Spain}

\author{P.~G. Kevrekidis}
\email[Email: ]{kevrekid@math.umass.edu}
\affiliation{Department of Mathematics and
Statistics, University of Massachusetts
Amherst, Amherst, MA 01003-4515, USA}
\affiliation{Mathematical Institute,
University of Oxford, Oxford, UK}

\author{N.~I. Karachalios}
\email[Email: ]{karan@aegean.gr}
\affiliation{Department of Mathematics,
University of the Aegean, Karlovassi,
83200 Samos, Greece}

\begin{abstract}
The Salerno model is a discrete variant of the celebrated nonlinear Schr\"odinger
(NLS) equation interpolating between the discrete NLS (DNLS) equation
and completely integrable Ablowitz-Ladik (AL) model by appropriately
tuning the relevant homotopy parameter. Although the AL model possesses an
explicit time-periodic solution known as the Kuznetsov-Ma (KM) breather,
the existence of time-periodic solutions away from the integrable limit
has not been studied as of yet. It is thus the purpose of this work to shed
light on the existence and stability of time-periodic solutions of the Salerno
model. In particular, we vary the homotopy parameter of the model by employing
a pseudo-arclength continuation algorithm where time-periodic solutions are
identified via fixed-point iterations. We show that the solutions transform
into time-periodic patterns featuring small, yet non-decaying far-field oscillations.
Remarkably, our numerical results support the existence of previously unknown
time-periodic solutions {\it even} at the integrable case whose stability is explored
by using Floquet theory. A continuation of these patterns towards the
DNLS limit is also discussed.
\end{abstract}

\date{\today}

\maketitle

\section{Introduction}
\label{sec:intro}

The study of rogue wave patterns has been a focal point of recent research 
in dispersive nonlinear wave systems~\cite{onorato,solli2,yan_rev,Akhmediev2016,Chen2017,Mihalache2017,MalomedMihalache2019}. 
Whether they may arise ``spontaneously out of nowhere and disappear without 
a trace''~\cite{wandt}, or generated gradually through energy transfer in 
multiple soliton collision, they are studied intensely in a diverse range of
settings~\cite{k2a,k2b,k2c,k2d}. Relevant studies have appeared from superfluid 
helium~\cite{He}, to plasmas~\cite{plasma} and from nonlinear optics~\cite{opt1,opt2,opt3,opt4,opt5,laser,genty}
to the, arguably, most  natural venue of water waves~\cite{hydro,hydro2,hydro3,hydro3b}.

In the discrete realm, there is far fewer studies. This is in a sense
relatively natural to expect. Much of what we know about rogue waves
is intimately connected to techniques stemming from integrable systems 
and the discrete setting of differential-difference equations is no 
exception. The most prototypical model associated with integrability
here is the so-called Ablowitz-Ladik (AL) model~\cite{AL1,AL2}.
In this setting, rogue waves in the form of the prototypical Peregrine 
soliton~\cite{H_Peregrine}, but also in the form of the Kuznetsov~\cite{kuz}, 
Ma~\cite{ma}, and Akhmediev~\cite{akh} breathers have been identified in 
the work of~\cite{akhm_AL}. The Peregrine soliton is a solution that is localized
in {\it both} space and time and it is a special member of a family extending 
between the spatially periodic Akhmediev solutions and the temporally-periodic 
Kuznetsov-Ma (KM) ones. Integrability has also provided higher order solutions~\cite{ohtayang}. 
However, beyond these findings there is very little. A question that is natural 
to ask is what of this structure persists in the more physically relevant (for 
settings such as waveguide arrays or Bose-Einstein condensates in optical lattices)
discrete nonlinear Schr{\"o}dinger (DNLS)~\cite{pgk_book} setting? A partial result
has stemmed from the statistical analysis of~\cite{tsironis}, which concluded that 
there is a large propensity towards freak waves near the integrable limit.

Over the past few years, we have attempted to address some of the relevant extensions 
of the understanding of rogue waves past the strict realm of integrable systems. On 
the one hand, some of the present authors have attempted to develop rogue-wave identifying
methods that go {\it beyond} the integrable realm~\cite{ward1,ward2} such as tracing 
these solutions as fixed points in space-time. On the other hand, a different subset 
of the present authors has developed techniques for understanding the stability of these 
states, considering the Floquet analysis of the KM waves and examining the Peregrine
states as natural limiting states thereof~\cite{jcm_pgk_1}. Dynamical studies of the 
evolution of generic initial data (including experimental ones)~\cite{bertola,suret,egc_10} 
are ongoing and such examples exist also in the case of discrete systems~\cite{egc_17}.

The purpose of the present work is multi-fold. We aim to use the so-called Salerno 
model~\cite{salerno} as a vehicle for going controllably beyond the integrable AL limit
and towards the physical realm of the DNLS limit. In principle, this is possible since the 
model itself involves a homotopy parameter, identified as $g$ hereafter (see, also Eq.~\eqref{salerno0}) 
which allows such a (continuous) deformation. The hope is that one can take this path ``all 
the way'', arriving at physically relevant solutions of the latter limit. Indeed, we will show
that in special cases an unprecedented possibility exists to carry through this program. However, 
generically our study will show that solutions in the form of KM solitons will exist; we seek 
specifically the latter because not only can we obtain the Peregrines as a special limit thereof, 
but also the stability is computationally tractable. We carry out the continuation of such states and
find surprisingly convoluted bifurcation diagrams. We also observe that these bifurcation diagrams
(over $g$) can be quite different for different breather frequencies and in special cases may
even continue to the DNLS limit. Importantly also, we find that the looping bifurcation diagrams
identified come back to ``intersect'' at a different point the integrable limit, presumably giving
rise to previously unknown, to the best of our understanding, integrable system solutions. Hence, 
our work may be of interest both to the applications of the system (bearing the first example of
such states ``reaching'' the DNLS limit), but also to the mathematical analysis of the integrable
system (providing previously unknown solutions with oscillatory tails in the AL limit).

Our presentation is structured as follows. In the next section, we offer the general framework 
of the model and the analytical results from the stability analysis of the plane wave solutions. 
Then in Section III, we examine our numerical findings for different frequencies of the periodic 
state $\omega$ and over the Salerno-model homotopic continuation strength $g$. Finally, in Section 
IV we summarize our findings and offer some possible directions for future study.

\section{Model and Theoretical setup}
\label{sec:model}
\subsection{Model and breather solutions}
The model that this work focuses on is the so-called Salerno model~\cite{salerno}
given by
\begin{equation}
i\dot{\Psi}_{n}+C\left(\Psi_{n+1}-2\Psi_{n}+\Psi_{n-1}\right)%
+g\left(\Psi_{n+1}+\Psi_{n-1}\right)|\Psi_{n}|^{2}+%
2\left(1-g\right)|\Psi_{n}|^{2}\Psi_{n}=0,
\label{salerno0}
\end{equation}
where $\Psi_{n}\coloneqq \Psi_{n}(t):\mathbb{Z}\times\mathbb{R}\to\mathbb{C}$ is
the complex wavefunction of the $n$th lattice site ($n\in\mathbb{Z}$) and overdot 
denotes differentiation with respect to time. The parameter $g\in[0,1]$ is the 
homotopy parameter whence the DNLS~\cite{pgk_book} and AL~\cite{AL1,AL2} models 
are obtained from Eq.~\eqref{salerno0} for values of $g=0$ and $g=1$, respectively. 
If $h$ stands for the distance between adjacent nodes, then $C=1/h^{2}$ in Eq.~\eqref{salerno0}.
Recall that the model is used due to its versatility. On the one hand, we can benefit 
from the knowledge of analytical solutions for the Peregrine soliton or the KM 
waveform at the AL limit, yet we can also attempt to connect (through continuations 
in $g$) these findings to the physically applicable  (non-integrable) DNLS limit.
Upon inserting the separation of variables ansatz
\begin{equation}
\Psi_{n}=\psi_{n}e^{2iq^{2}t}
\end{equation}
into Eq.~\eqref{salerno0}, we arrive at
\begin{equation}
i\dot{\psi}_{n}+C\left(\psi_{n+1}-2\psi_{n}+\psi_{n-1}\right)%
+g\left(\psi_{n+1}+\psi_{n-1}\right)|\psi_{n}|^{2}+%
2\left[\left(1-g\right)|\psi_{n}|^{2}-q^{2}\right]\psi_{n}=0,
\label{salerno}
\end{equation}
where the parameter $q$ fixes the background amplitude.

At the AL limit, i.e., when $g=1$ and $C=1$, Eq.~\eqref{salerno} possesses a time-periodic
solution, the Kuznetsov-Ma breather~\cite{akhm_AL} given explicitly by
\begin{equation}
\psi_{n}(t)=q\frac{\cos{\left(\omega t+i\theta\right)+G\cosh{\left(r n\right)}}}%
{\cos{\left(\omega t\right)+G\cosh{\left(r n\right)}}},
\label{km_exact}
\end{equation}
with frequency $\omega$ (related to the period of the solution via $T=2\pi/\omega$), 
$\theta=-\arcsinh{\left(\omega\right)}$, $r=\arccosh{\left(\left[2+\cosh{(\theta)}\right]/3\right)}$ 
and $G=-\omega/\left(\sqrt{3}\sinh{\left(r\right)}\right)$.

Away from the integrable AL limit, explicit analytical solutions are no longer 
available. Thus, we identify time-periodic solutions of period $T$ by considering 
a temporal discretization in terms of the Fourier series expansion:
\begin{equation}
\psi_{n}(t)=\sum_{m=-\infty}^{\infty}\phi_{n,m}e^{im\omega t},
\label{fourier_decomp}
\end{equation}
where $\phi_{n,m}$ are the Fourier coefficients. Upon plugging Eq.~\eqref{fourier_decomp} 
into Eq.~\eqref{salerno}, we arrive at a set of algebraic equations:
\begin{equation}
\begin{split}
-&(m\omega+2q^2)\phi_{n,m}+C\left(\phi_{n+1,m}-2\phi_{n,m}+\phi_{n-1,m}\right)-2q^2 \\
+&g\sum_{m'}\sum_{m''}\lbrace \phi_{n,m'}\phi^{\ast}_{n,m''}\left(\phi_{n-1,m-m'+m''}+\phi_{n+1,m-m'+m''}\right) \\
+&2(1-g)\phi_{n,m'}\phi^*_{n,m''}\phi_{n-1,m-m'+m''}\rbrace=0.
\label{algebraic_eqs}
\end{split}
\end{equation}
This system is solved by a fixed-point iteration method, e.g., the Newton-Raphson method. 
To this aim, one must firstly truncate the infinite spatial lattice into a finite one to 
render the system tractable. Here we choose $n\in\left[-N/2,N/2\right]$, $N\in2\mathbb{Z}^{+}$ 
supplemented with periodic boundary conditions $\psi_{-N/2}(t)=\psi_{N/2}(t),\,\forall t$. 
This way, the total number of nodes is $N+1$. On the other hand, one must truncate the 
infinite (temporal) Fourier series of Eq.~\eqref{fourier_decomp}. We use $|m|\leq41$ in 
our computations presented below. For given $\omega$, $C$ and $g$, we identify time-periodic 
solutions with high accuracy by imposing a strict tolerance criterion on the norm of successive 
iterates which is within $10^{-11}$. It should be noted in passing that upon convergence of
the Newton-Raphson method, we construct the time-periodic solution by means of Eq.~\eqref{fourier_decomp}~\cite{jcm_pgk_1}. 
The resulting, by construction, time-periodic solution can be used as an initial condition 
(at $t=0$) for a time-stepping scheme to examine its dynamical evolution. Additionally, it can
also be used in a Floquet analysis-based stability computation to assess the spectral stability 
of the solution as we now discuss.
\subsection{Modulational instability and Floquet analysis}
The modulational instability (MI) of the asymptotic state (constant background) 
$\Psi_{n}=q$ as $|n|\to\infty$ of Eq.~\eqref{salerno0} (see, the seminal works of~\cite{kivshar_mi,abdullaev_mi} 
as well as the recent work of~\cite{ndzana_mohamadou} on the subject) is investigated
by considering the plane wave solution
\begin{equation}
\Psi_{n}=q e^{i\left(kn-\omega t\right)}, \quad k=2\pi M\left(\frac{\sqrt{C}}{N}\right), \quad M\in\mathbb{Z},
\label{cw}
\end{equation}
with frequency $\omega$ and wavenumber $k$. If we insert Eq.~\eqref{cw} to Eq.~\eqref{salerno0} 
we obtain the following dispersion relation:
\begin{equation}
\omega=4\left(C+gq^{2}\right)\sin^{2}{\left(\frac{k}{2}\right)}-2q^{2}.
\label{disp_0}
\end{equation}
We can now explore the linear stability of the plane wave solution of Eq.~\eqref{cw} by 
introducing the ansatz
\begin{equation}
\widetilde{\Psi}_{n}=\left[q+\varepsilon\left(a+ib\right)%
e^{i\left(Qn-\omega t\right)}\right]%
e^{i\left(kn-\omega t\right)},
\quad \varepsilon\ll 1, \quad Q=2\pi M'\left(\frac{\sqrt{C}}{N}\right), \quad M'\in\mathbb{Z},
\label{cw_pertr}
\end{equation}
where both $a$ and $b$ are time-dependent, real-valued functions, and $Q$ and $\Omega$ 
correspond to the wavenumber and frequency of the perturbation, respectively. Upon 
plugging Eq.~\eqref{cw_pertr} into Eq.~\eqref{salerno0}, we obtain at order 
$\mathcal{O}(\varepsilon)$ the MI dispersion relation given by
\begin{equation}
\Omega^{2}=2\left(gq^{2}+C\right)\left[\cos{\left(Q+k\right)}-\cos{(k)}\right]%
\left[2\left(gq^{2}+C\right)\cos{\left(Q+k\right)}+2\left(gq^{2}-C\right)\cos{(k)}%
+4\left(1-g\right)q^{2}\right].
\label{mi_freq}
\end{equation}
If this condition is satisfied, yielding real frequencies $\Omega$ for
a given perturbation wavenumber $Q$, then the relevant wavenumber
is stable. On the other hand, the existence of $Q$'s associated with
complex $\Omega$ leads to dynamical instability of the background.

The stability of time-periodic solutions with period $T$ identified via fixed-point 
iterations (see, Sec.~\ref{sec:num_results}) and denoted as $\psi_{n}^{0}$ is examined
by considering the perturbation ansatz
\begin{equation}
\widetilde{\psi}_{n}=\psi_{n}^{0}+\varepsilon\,\xi_{n}(t), \quad \varepsilon\ll 1,
\label{petru_ansatz}
\end{equation}
where $\xi_{n}\in\mathbb{C}$ is the perturbation imposed at the $n$th site of the lattice. 
Then, we insert Eq.~\eqref{petru_ansatz} into Eq.~\eqref{salerno} and obtain at order
${\mathcal{O}}(\varepsilon)$ the governing equation for the perturbation $\xi_{n}$:
\begin{align}
i\dot{\xi}_{n}=&-C\left(\xi_{n+1}-2\xi_{n}+\xi_{n-1}\right)-%
2g\left(\psi_{n+1}^{0}+\psi_{n-1}^{0}\right)\textrm{Re}\left(\xi_{n}(\psi_{n}^{0})^{\ast}\right)-%
g\left(\xi_{n+1}+\xi_{n-1}\right)|\psi_{n}^{0}|^{2}\nonumber\\ 
&-2\left(1-g\right)%
\left[2|\psi_{n}^{0}|^{2}\xi_{n}+\xi_{n}^{\ast}(\psi_{n}^{0})^{2}\right]+2q^{2}\xi_{n}.
\label{var_eqs}
\end{align}
Then, the eigenvalues $\lambda$ of the so-called monodromy matrix $\mathcal{M}$ stemming
from:
\begin{equation}
\left[\begin{array}{c}
  \textrm{Re}(\xi_{n}(T)) \\ \textrm{Im}(\xi_{n}(T)) \\
  \end{array}\right]
  =\mathcal{M}
  \left[\begin{array}{c}
  \textrm{Re}(\xi_{n}(0)) \\ \textrm{Im}(\xi_{n}(0)) \\
  \end{array}\right].
\label{variational_prob}
\end{equation}
determine the stability trait of of a time-periodic solution $\psi_{n}^{0}$. 
Those eigenvalues are the so-called Floquet multipliers. In particular, as 
the system is symplectic and Hamiltonian, a solution is deemed neutrally stable 
if all the Floquet multipliers $\lambda=\lambda_{r}+i\lambda_{i}$ of $\mathcal{M}$ 
lie on the unit circle. If $|\lambda|>1$, then two types of instabilities can arise. 
If the multipliers arise in real pairs that are away from the unit circle, the 
instability is considered exponential due to the exponential growth of the associated 
perturbations. On the contrary, if the multipliers arise in complex quartets (inside 
and outside the unit circle), then the instability is deemed oscillatory. We conclude
this section by mentioning in passing that the eigenfrequency of the perturbation $\Omega$
connects with the Floquet multipliers $\lambda$ via
\begin{equation}
\lambda=e^{i\Omega T}.
\label{floq_to_sigma}
\end{equation}

\section{Numerical Results}
\label{sec:num_results}

\begin{figure}[!pbt]
\begin{center}
\includegraphics[height=.18\textheight, angle =0]{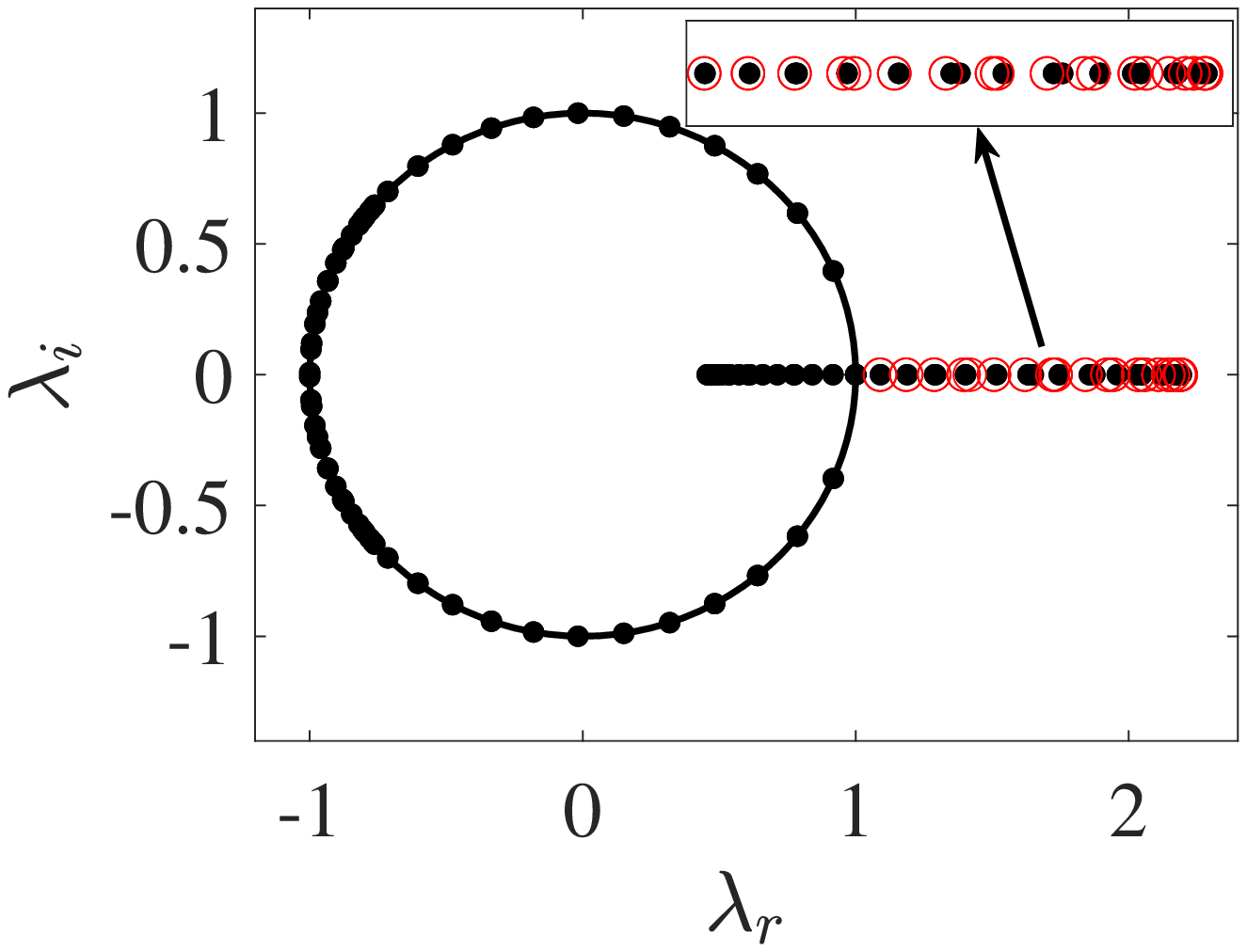}
\includegraphics[height=.18\textheight, angle =0]{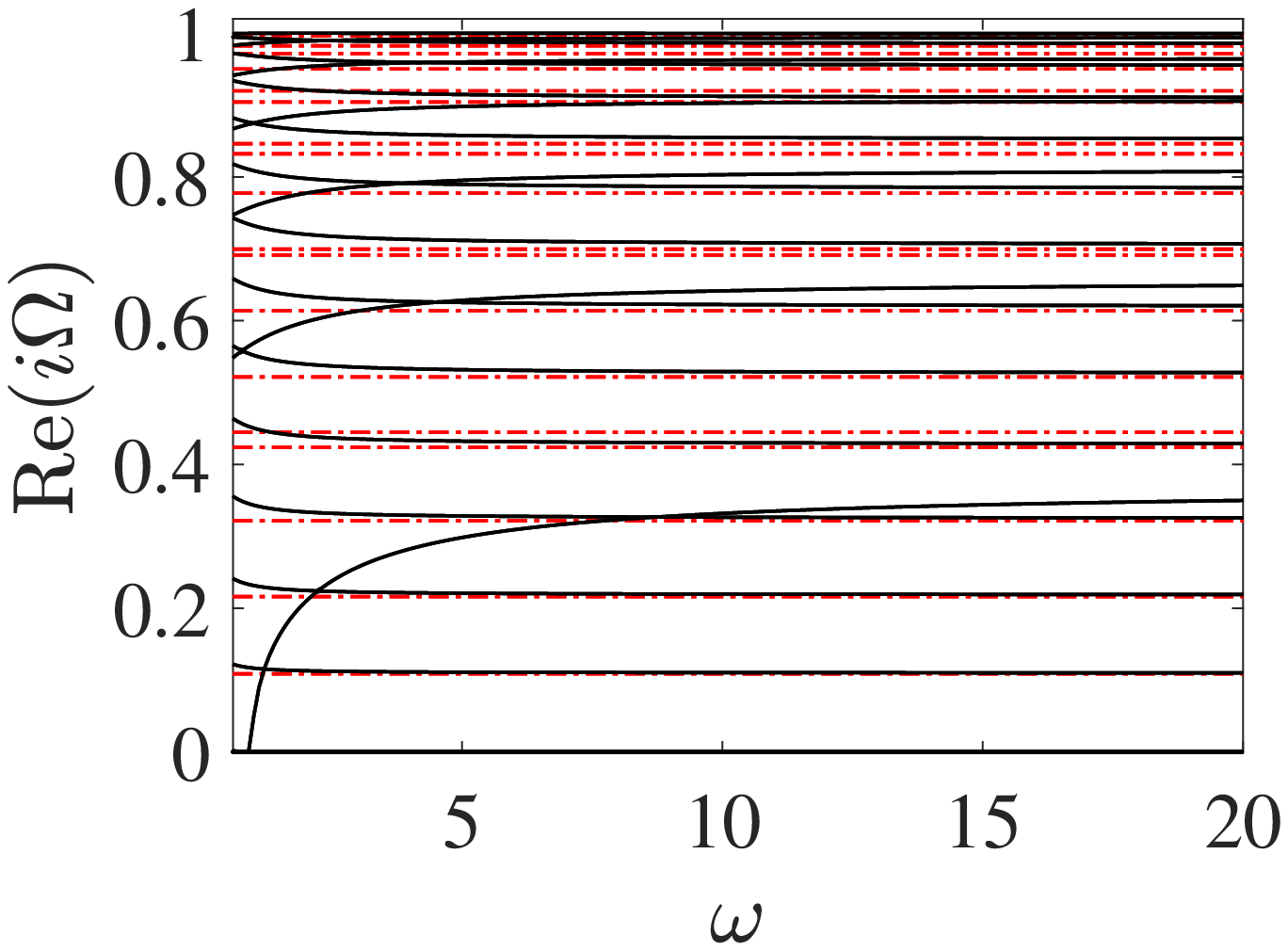}
\end{center}
\caption{(Color online) The left panel presents the spectrum of 
a(n unstable) time-periodic solution with $\omega=8$ (filled black 
circles). Note that the theoretically predicted unstable modes via
Eq.~\eqref{mi_freq} are shown with red open circles therein. The right 
panel demonstrates a comparison between the Floquet exponents 
$\textrm{Re}(i\Omega)$ of time-periodic solutions (solid black lines) 
and the ones of the MI analysis (dashed-dotted red lines) of 
Eq.~\eqref{mi_freq} over $\omega$. Note that both panels correspond
to the case with $g=1$ (as well as $C=1$ and $q=1/\sqrt{2}$).
}
\label{fig1}
\end{figure}

The availability of the exact solution~\eqref{km_exact} for the case with 
$g=1$, i.e., the AL limit allows us to not only benchmark our numerical 
methods but also to compute the Floquet multipliers directly from Eq.~%
\eqref{variational_prob}. Hereafter, we consider a lattice with $N=100$ 
sites and set $C=1$ and $q=1/\sqrt{2}$ in Eqs.~\eqref{algebraic_eqs} and~\eqref{var_eqs}
(all of our numerical results discussed in this section were obtained for 
these choices). At first, we compute the Floquet multipliers from Eq.~\eqref{variational_prob} 
(for $g=1$) by using the initial-value-problem (IVP) solver \verb|DOP853|~\cite{Hairer}
with (relative and absolute) tolerances $10^{-11}$. Although a non-stiff 
and explicit IVP solver, \verb|DOP853| can perform a time step-size adaptation 
(to satisfy the user-specified tolerance criteria) for stiff regions by reducing
the time step-size. Indicatively, Fig.~\ref{fig1} summarizes our findings on 
the Floquet multipliers for the exact solution~\eqref{km_exact}. The left 
panel of the figure presents the Floquet multipliers of the exact time-periodic 
solution with $\omega=8$ (filled black circles) together with the theoretically 
predicted unstable modes from Eq.~\eqref{mi_freq} via Eq.~\eqref{floq_to_sigma}
(open red circles). As far as the KM breather is concerned, the presence of Floquet
multipliers with $|\lambda|>1$ render the solution unstable. In addition, it is evident 
from this panel that a subset of the unstable modes of the KM breather coincides 
with those of the modulationally unstable background. However, there exist other
unstable eigenmodes that deviate from the latter due to the presence of the 
localized solution perturbing the modes of the background. This is natural to 
expect given the modulational instability of the background (analyzed in the 
previous section) on top of which the periodic solution ``lives''. The 
right panel of Fig.~\ref{fig1} complements our analysis of the exact KM 
breather and demonstrates the dependence of its unstable Floquet exponents 
$\textrm{Re}(i\Omega)$ on $\omega$ (solid black line) together with the 
respective results of the MI analysis (dashed-dotted red line). It can be
discerned from this panel that the Floquet exponents of the KM breather 
approach the asymptotic values theoretically predicted by Eq.~\eqref{mi_freq}.

We now wish to explore the existence and stability of time-periodic solutions
over the relevant parameter space by going beyond the well-defined and analytically 
tractable case of the integrable limit of $g=1$. We thus vary the parameter 
$g$ (which is used as our bifurcation parameter) by employing a pseudo-arclength 
continuation algorithm~\cite{doedel_I,ps_method}, and consider different fixed 
values of the period $T$. It should be noted that the pseudo-arclength continuation 
algorithm is capable of passing through turning points (where the Jacobian of the 
system of equations is singular and thus non-invertible) and tracing (connected) 
branches of solutions. The numerical results reported in this work were obtained by
using a fixed and relatively small arclength step-size of $10^{-2}$ in order to 
prevent the predictor-corrector step from converging to a solution of a nearby branch.
As per the direct numerical simulations reported below, we use again the \verb|DOP853|~\cite{Hairer} 
method for advancing Eq.~\eqref{salerno0} forward in time (and with the same tolerance
criteria as before).
\begin{figure}[!pt]
\begin{center}
\includegraphics[height=.18\textheight, angle =0]{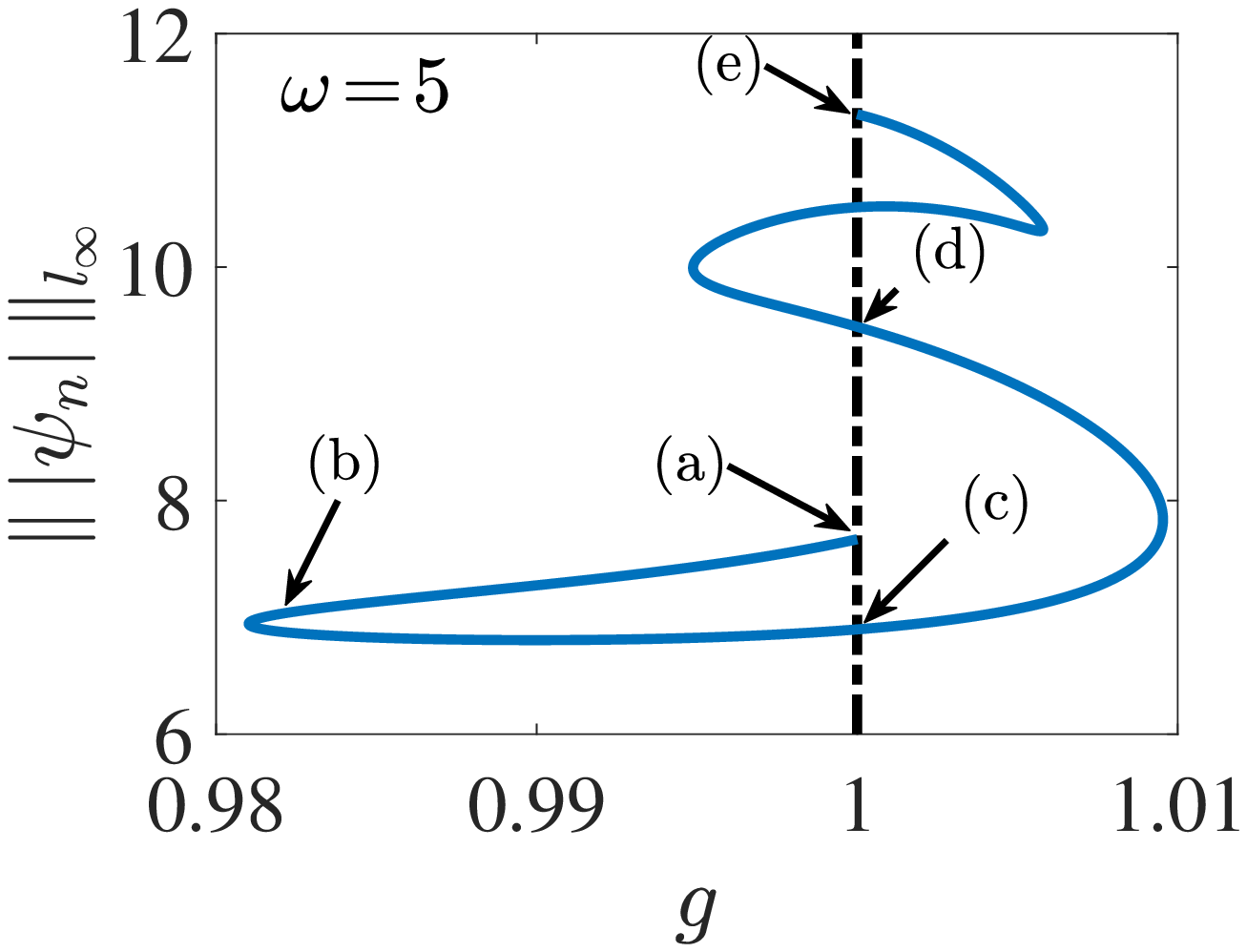}
\includegraphics[height=.18\textheight, angle =0]{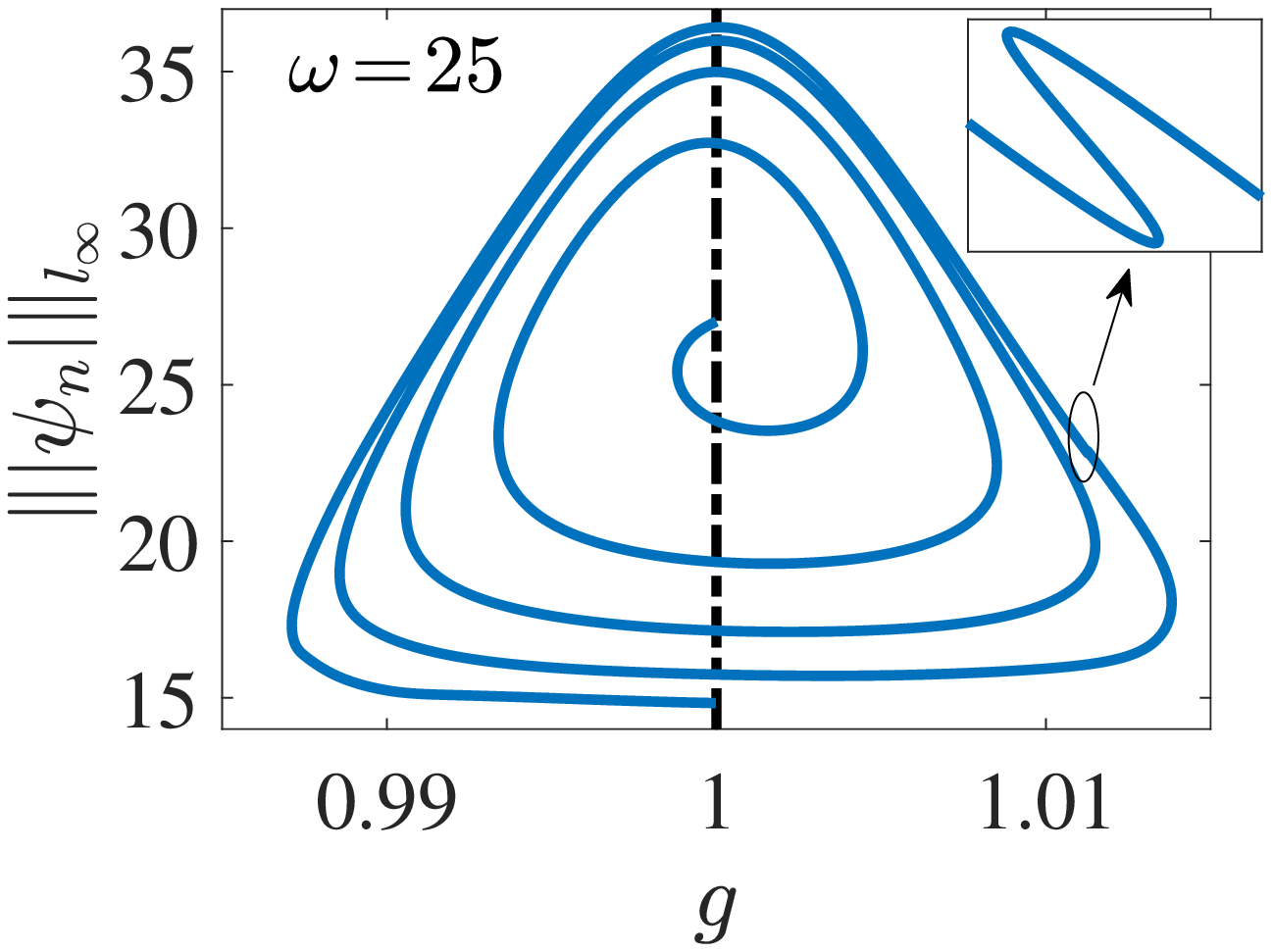}
\includegraphics[height=.18\textheight, angle =0]{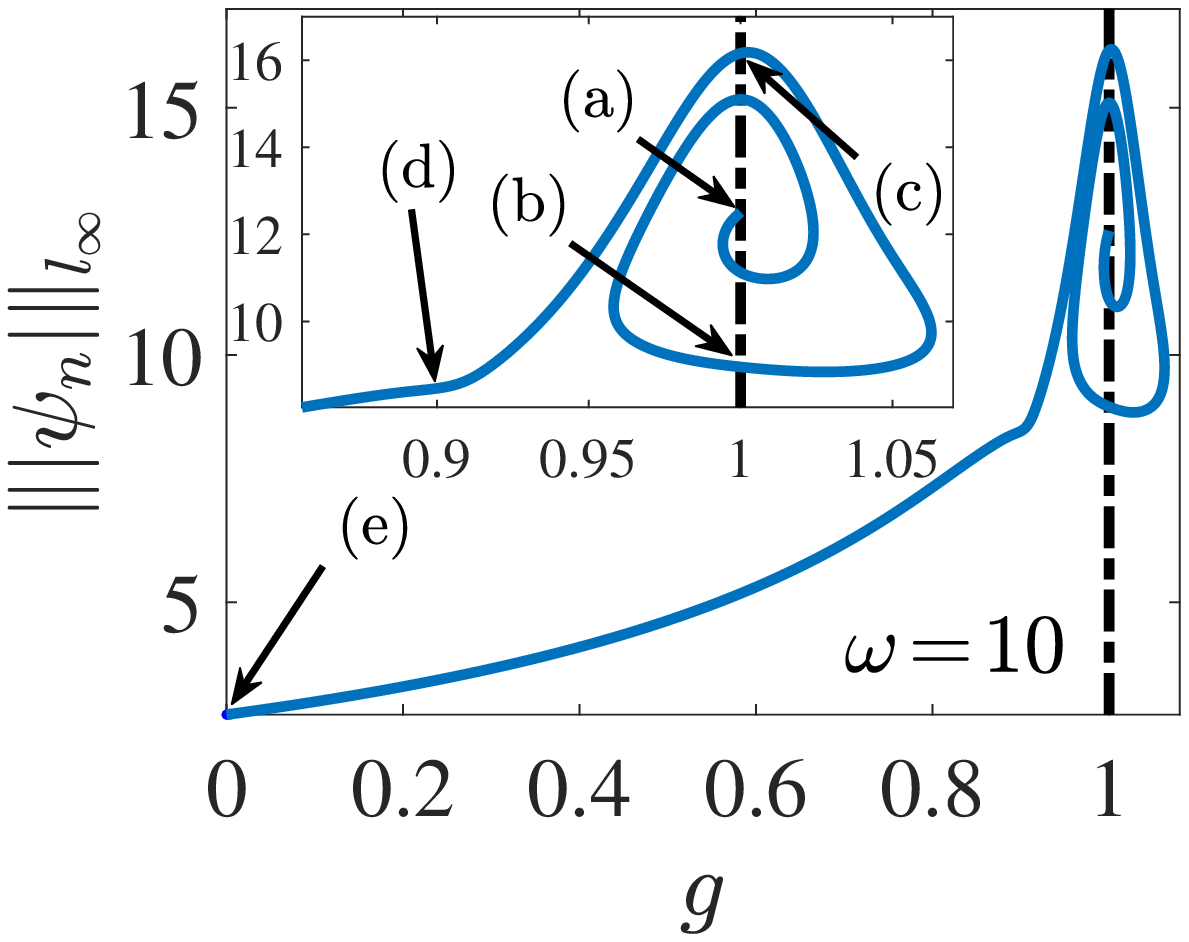}
\end{center}
\caption{(Color online) The $l_{\infty}$-norm of $|\psi_{n}|$
as a function of $g$ corresponding to time-periodic solutions
for values of $\omega=5$, $\omega=25$ and $\omega=10$, respectively.  
The labels (a)-(e) in the left and right panels are associated 
with Figs.~\ref{fig3} and \ref{fig4} that follow.
}
\label{fig2}
\end{figure}

\begin{figure}[!pht]
\begin{center}
\includegraphics[height=.18\textheight, angle =0]{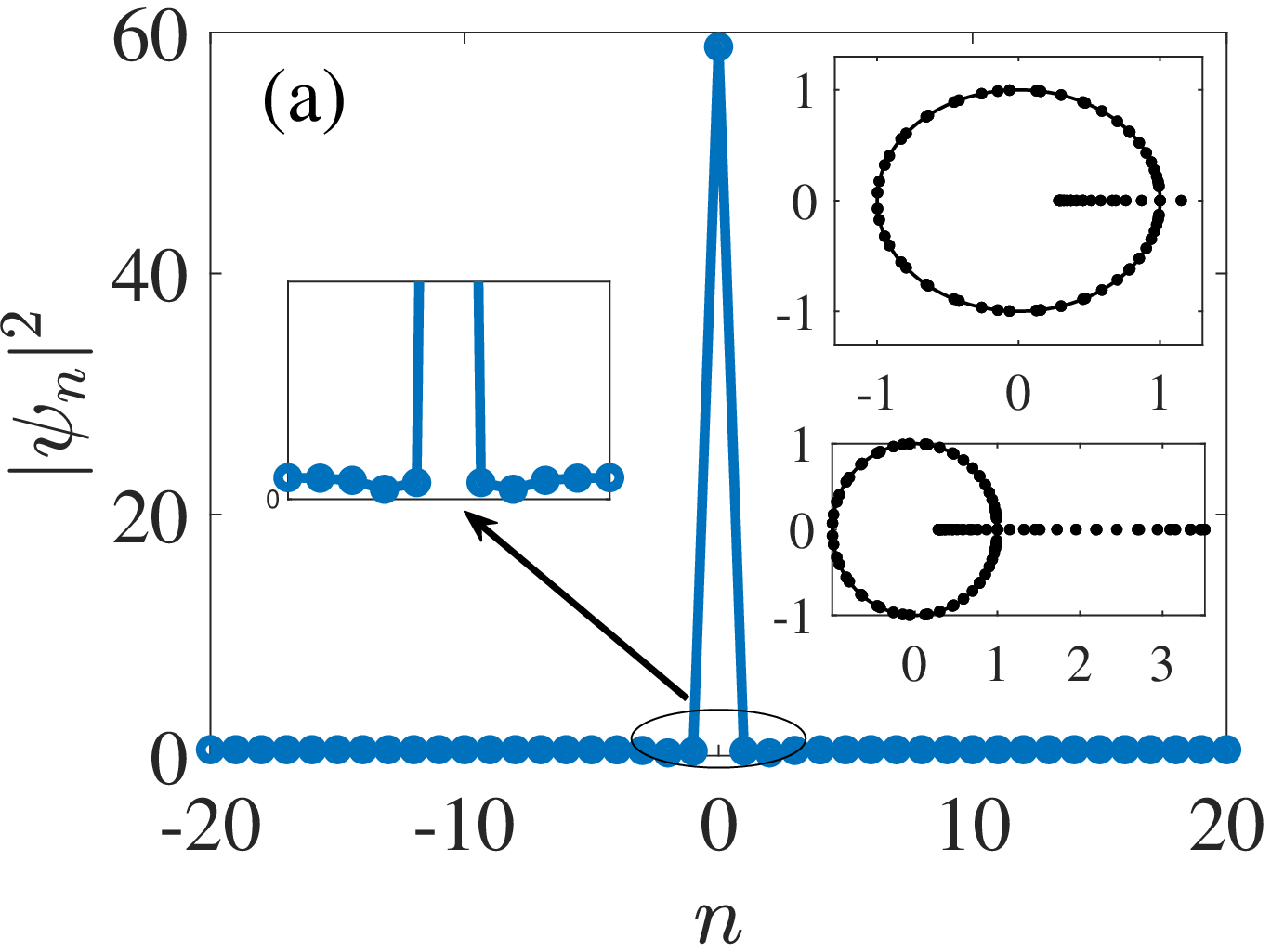}
\includegraphics[height=.18\textheight, angle =0]{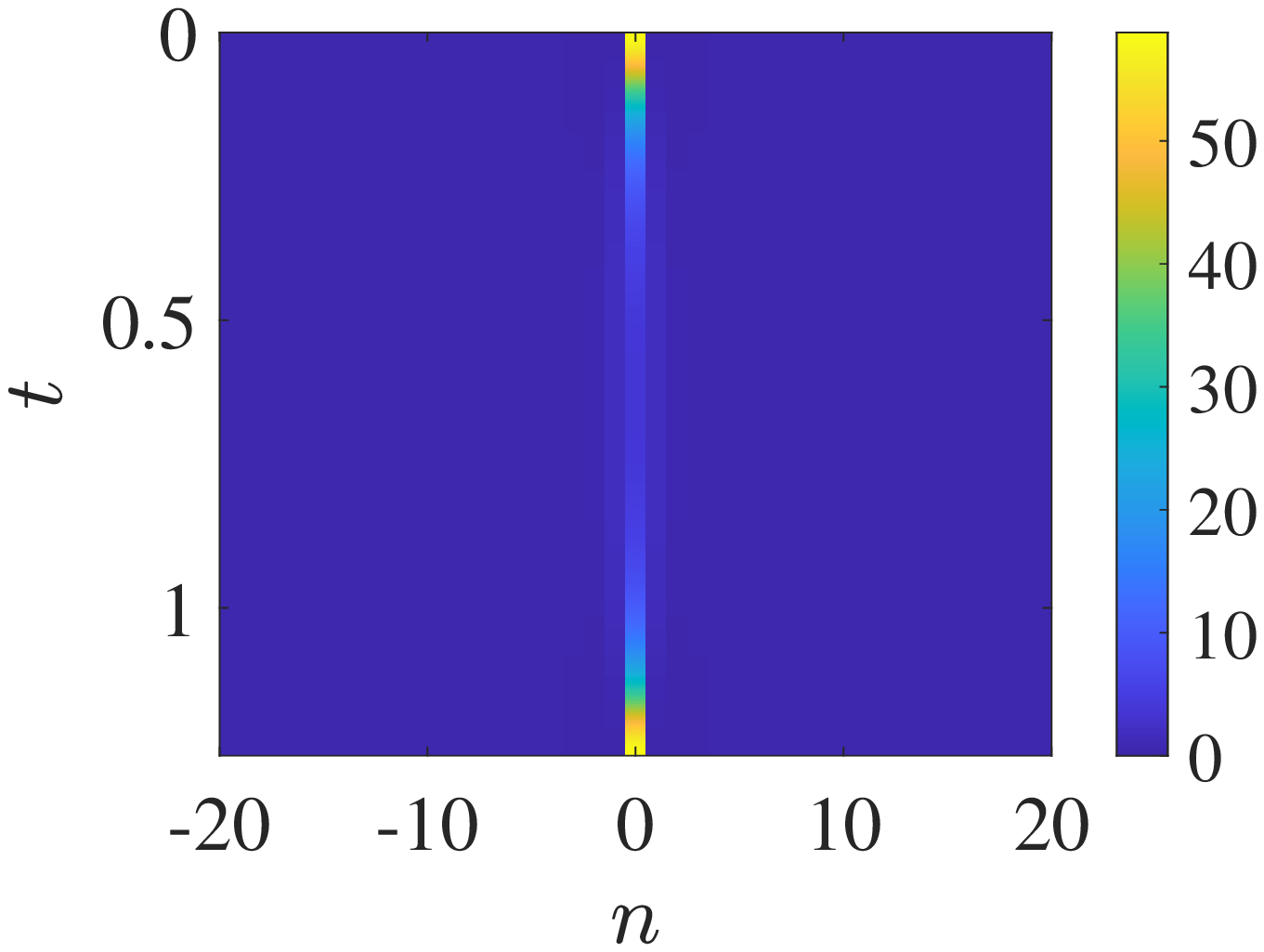}\\
\includegraphics[height=.18\textheight, angle =0]{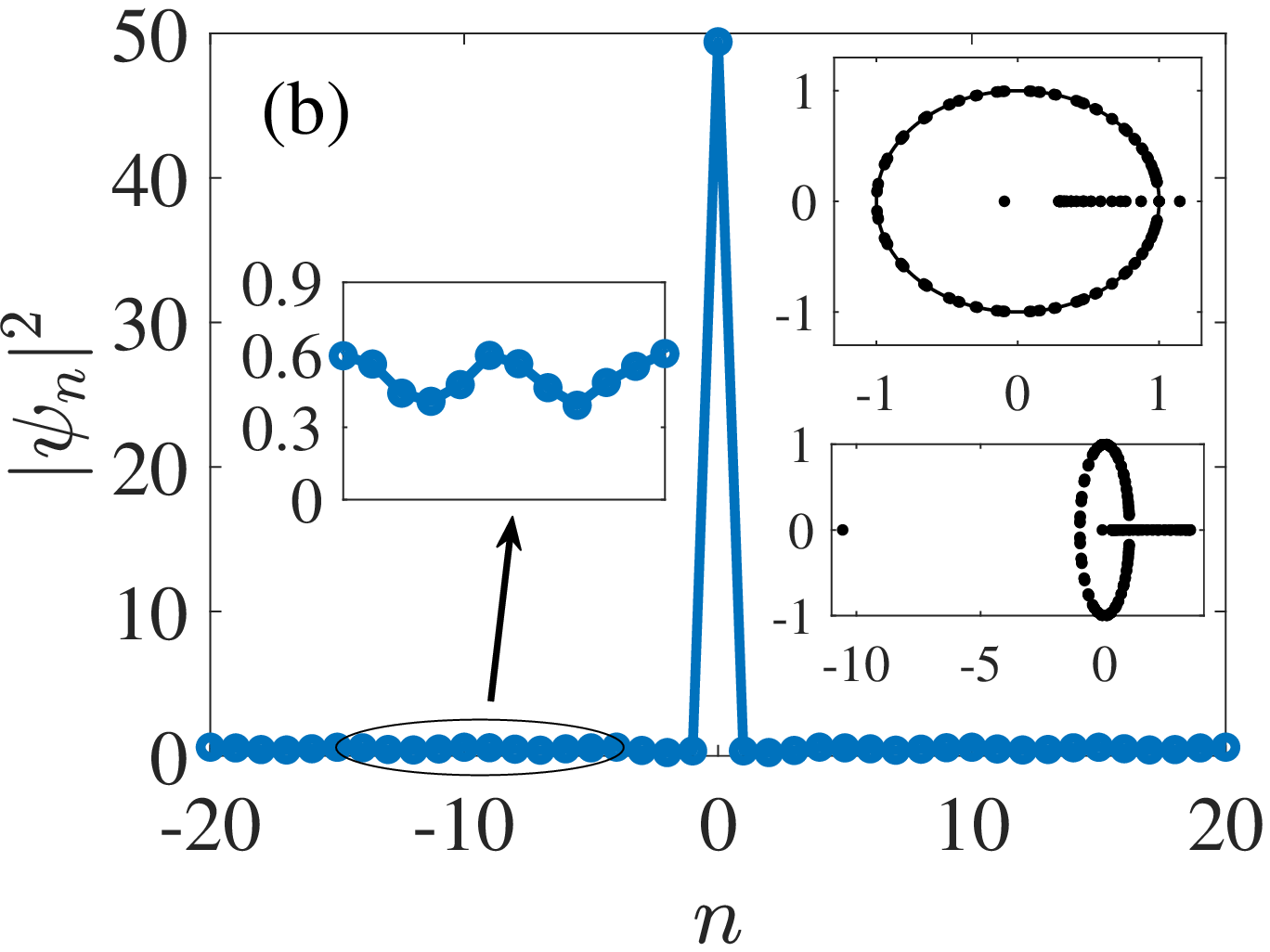}
\includegraphics[height=.18\textheight, angle =0]{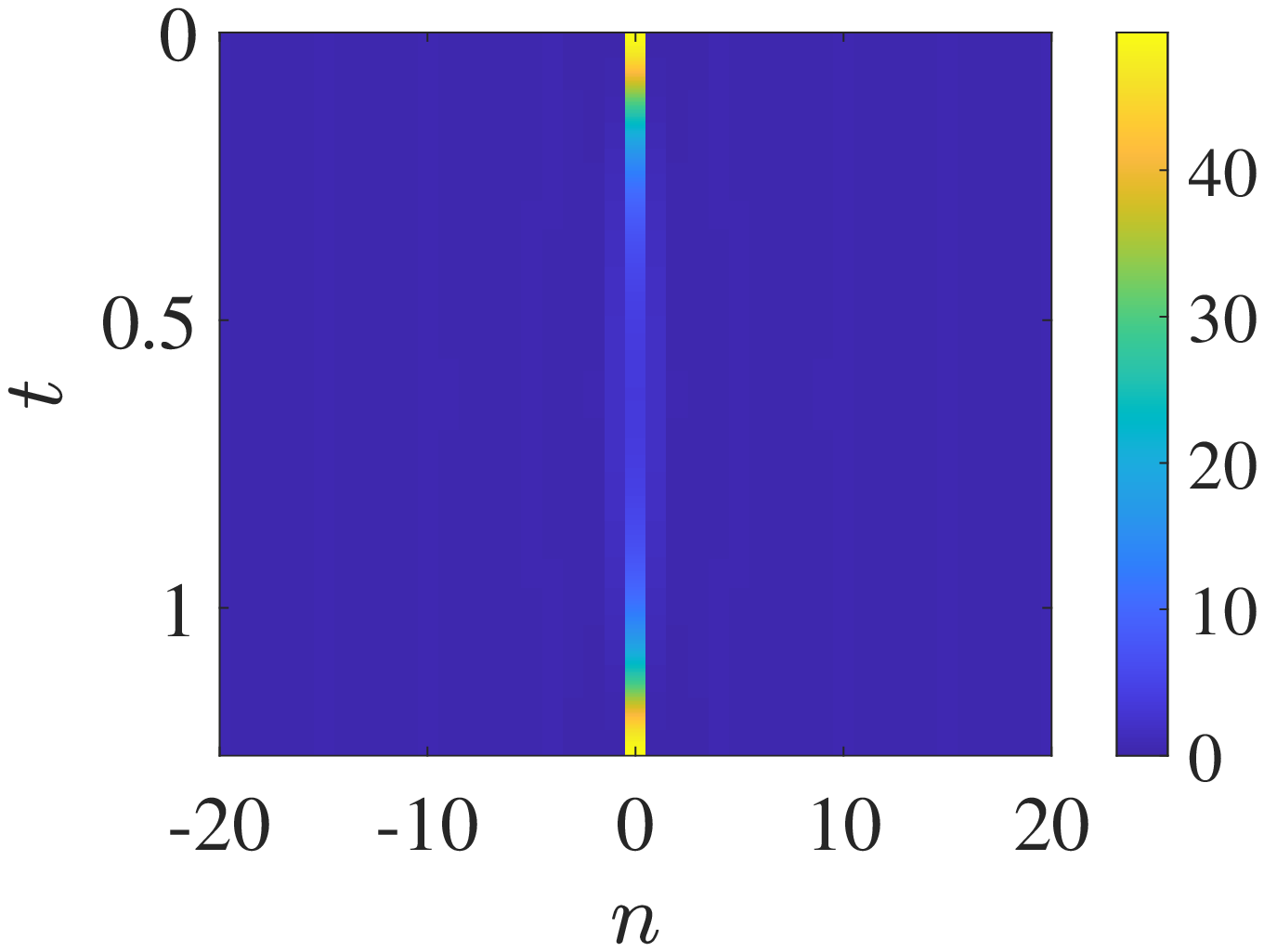}\\
\includegraphics[height=.18\textheight, angle =0]{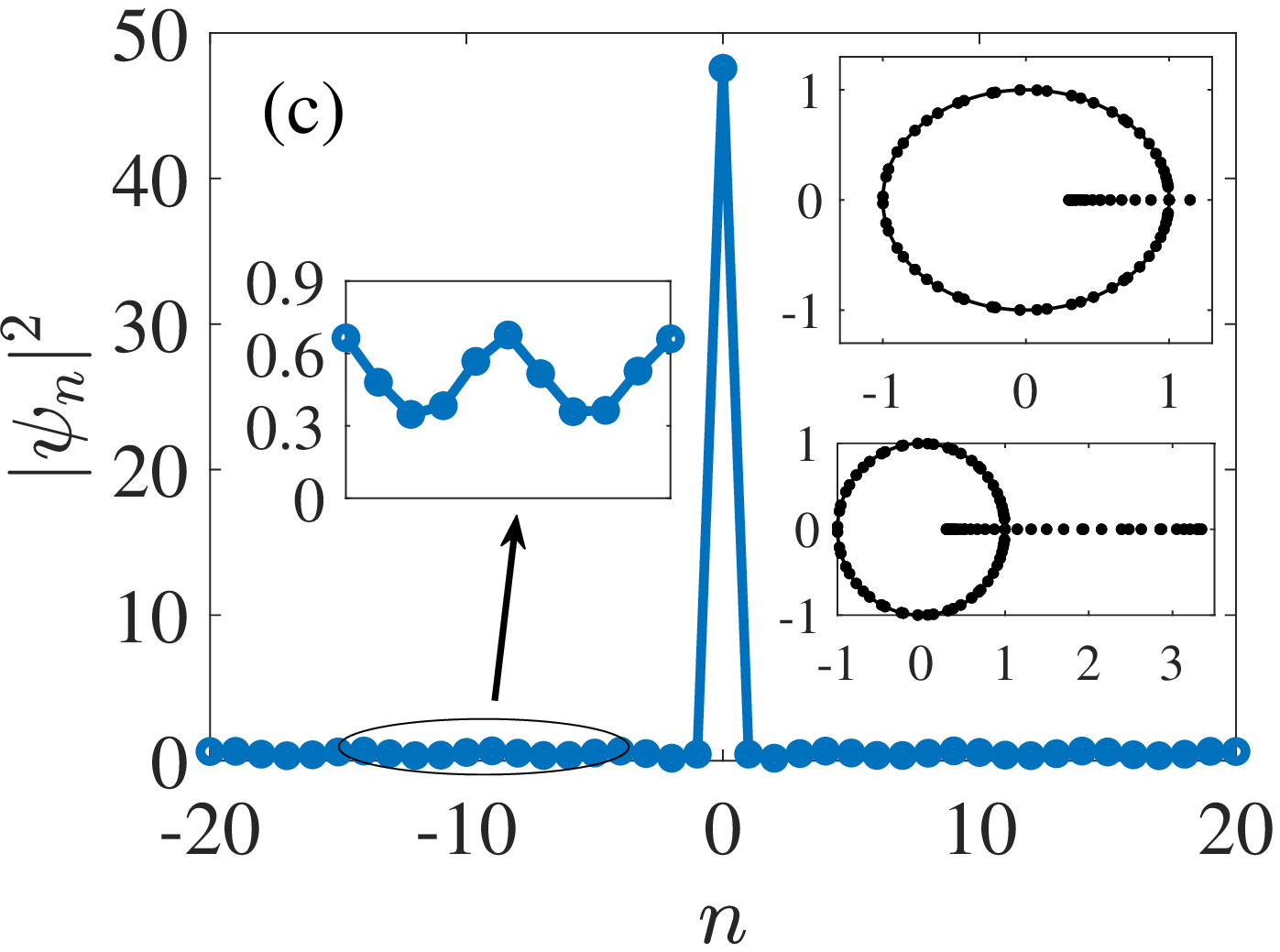}
\includegraphics[height=.18\textheight, angle =0]{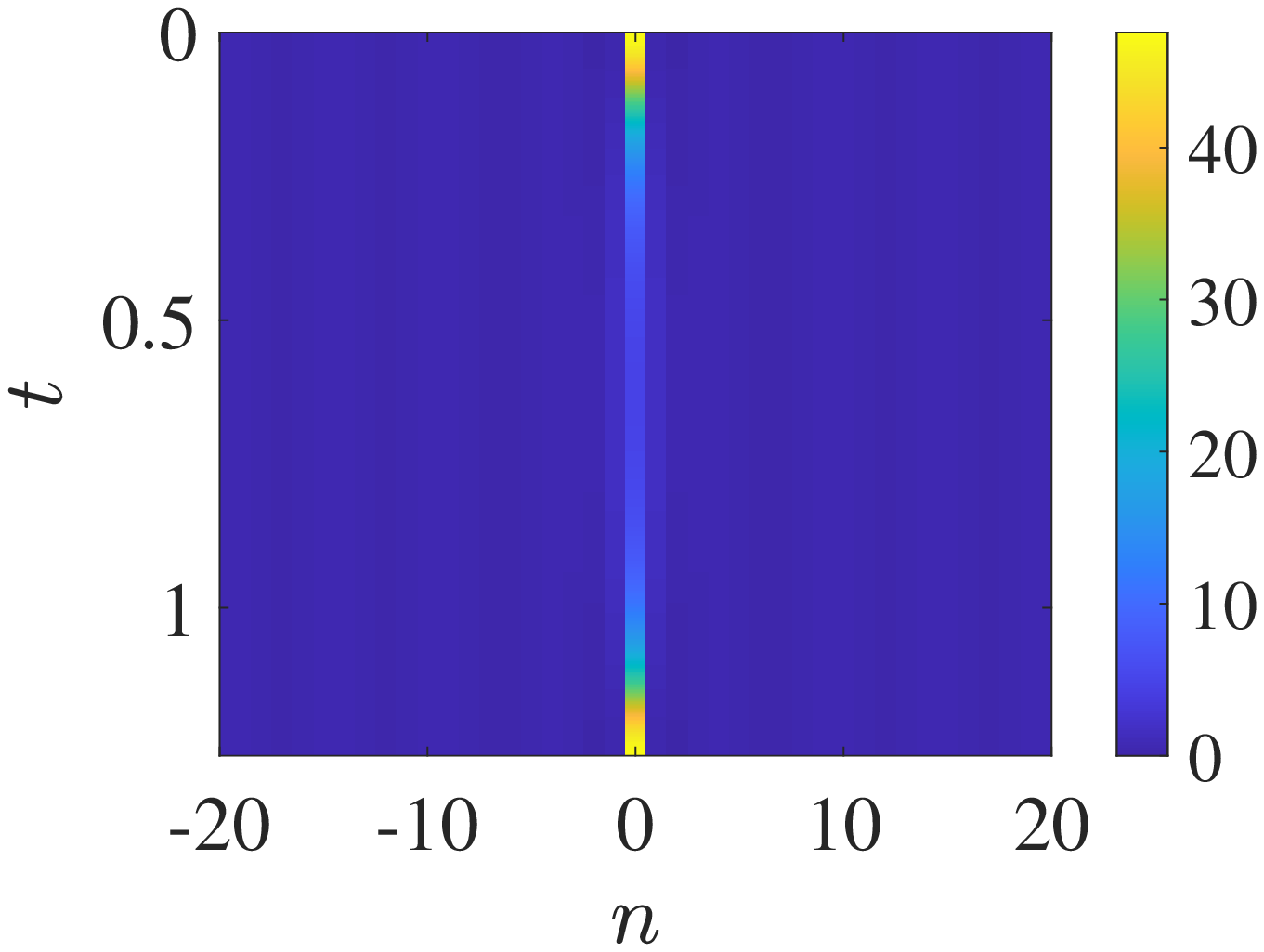}\\
\includegraphics[height=.18\textheight, angle =0]{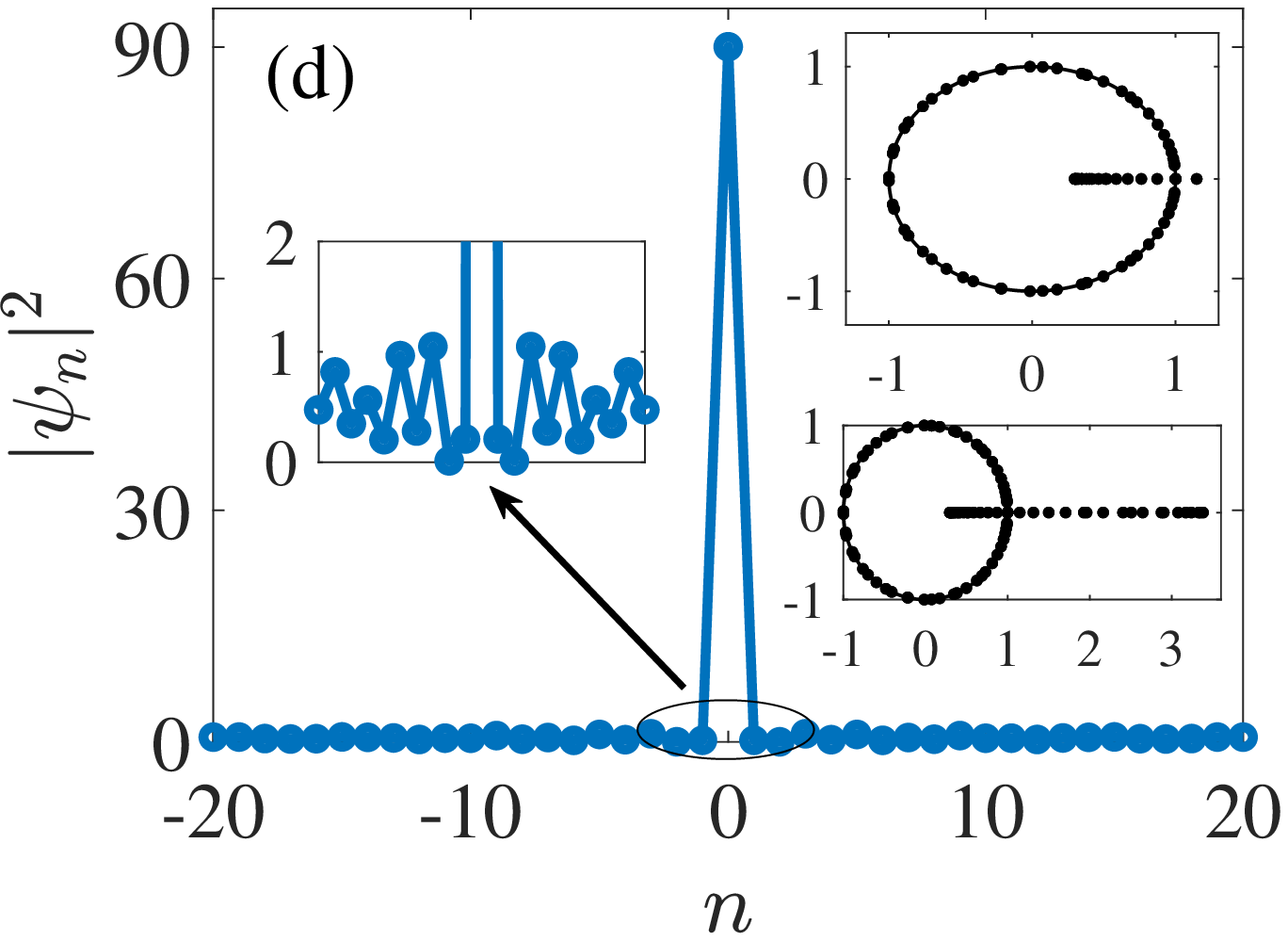}
\includegraphics[height=.18\textheight, angle =0]{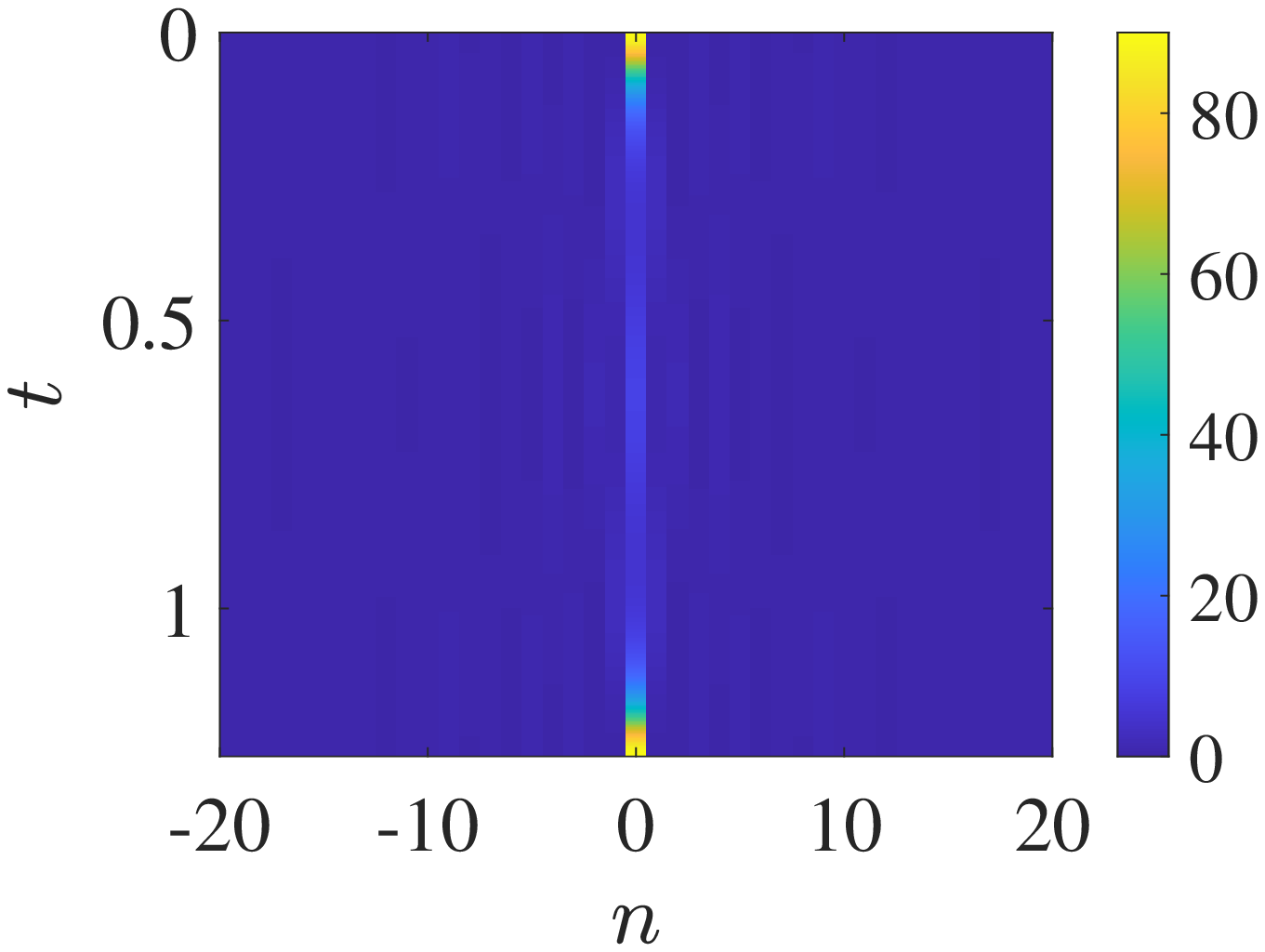}\\
\includegraphics[height=.18\textheight, angle =0]{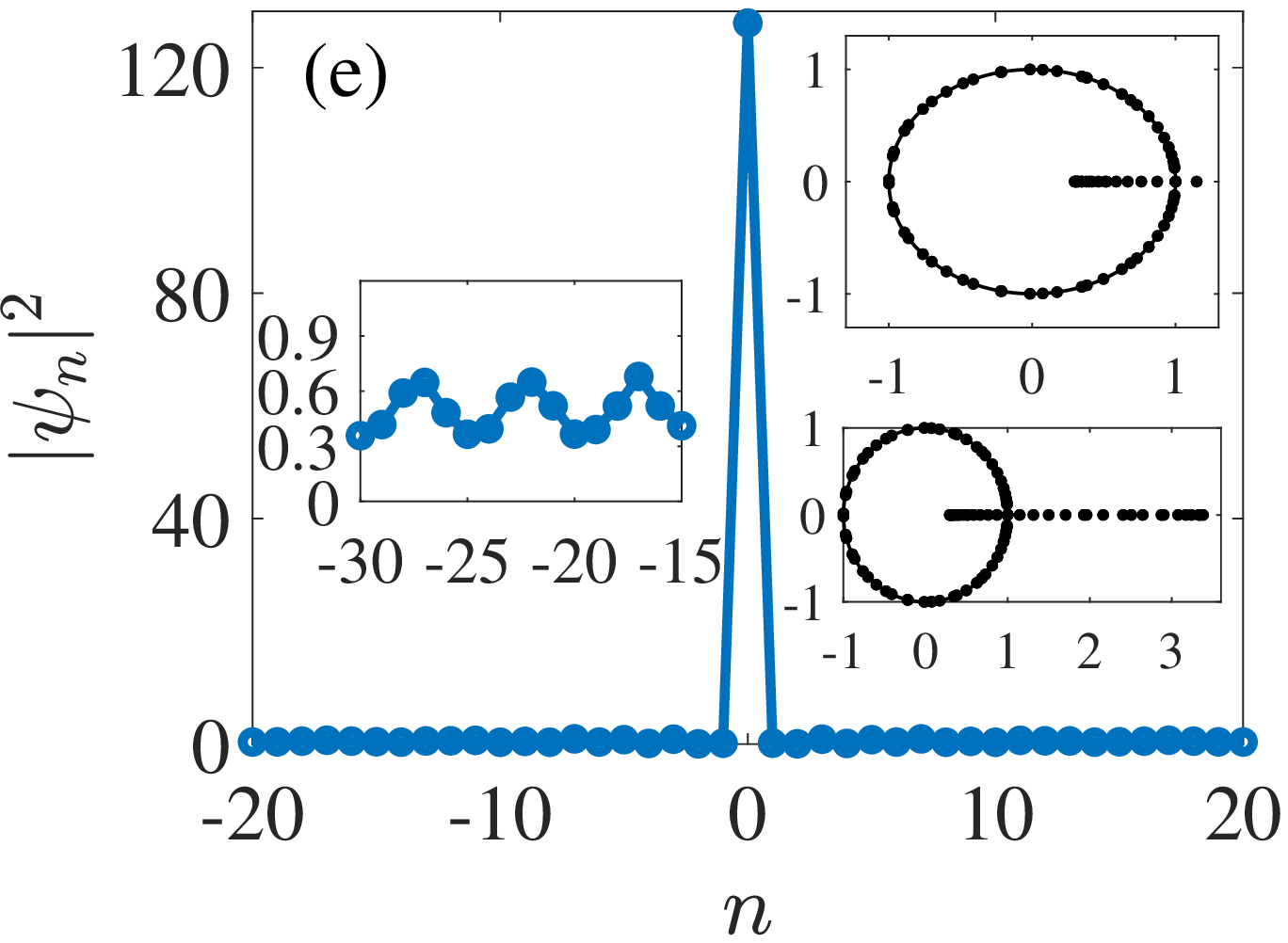}
\includegraphics[height=.18\textheight, angle =0]{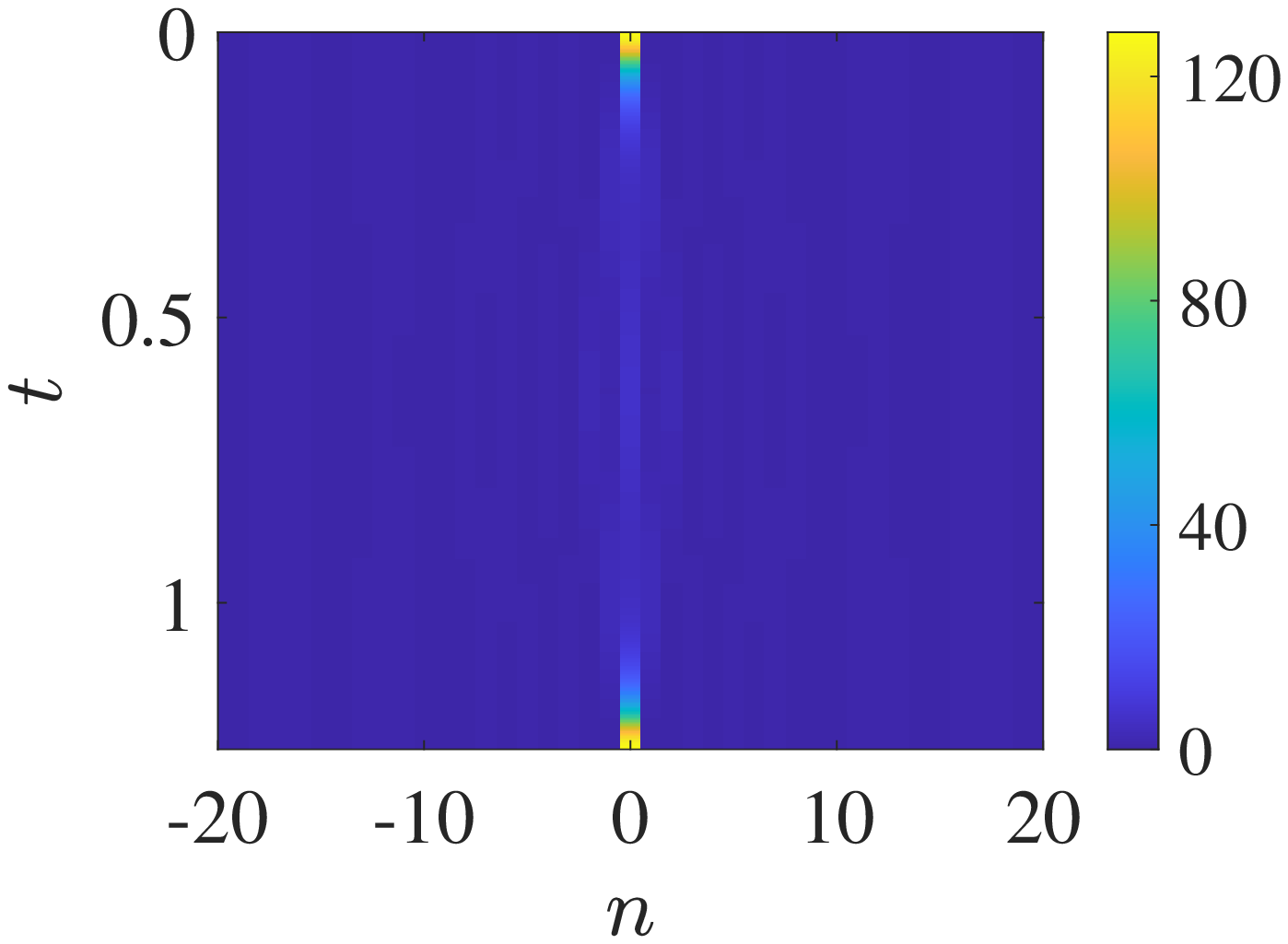}
\end{center}
\caption{(Color online) Numerical results on time-periodic
solutions of the Salerno model~\eqref{salerno} for $\omega=5$
associated with the bifurcation curve of the left panel of 
Fig.~\ref{fig2} (see, the labels therein). The left column 
corresponds to the spatial distribution of $|\psi_{n}|^{2}$ 
for values of (a), (c)-(e) $g=1$ and (b) $g=0.982$, respectively. 
The insets shown on the left and right of each panel provide 
close-ups of the profiles as well as Floquet spectra. The right 
column presents the spatio-temporal evolution of $|\psi_{n}|^{2}$ 
of the profiles shown in the left column for one period ($T=2\pi/5$).
}
\label{fig3}
\end{figure}

\begin{figure}[!pht]
\begin{center}
\includegraphics[height=.18\textheight, angle =0]{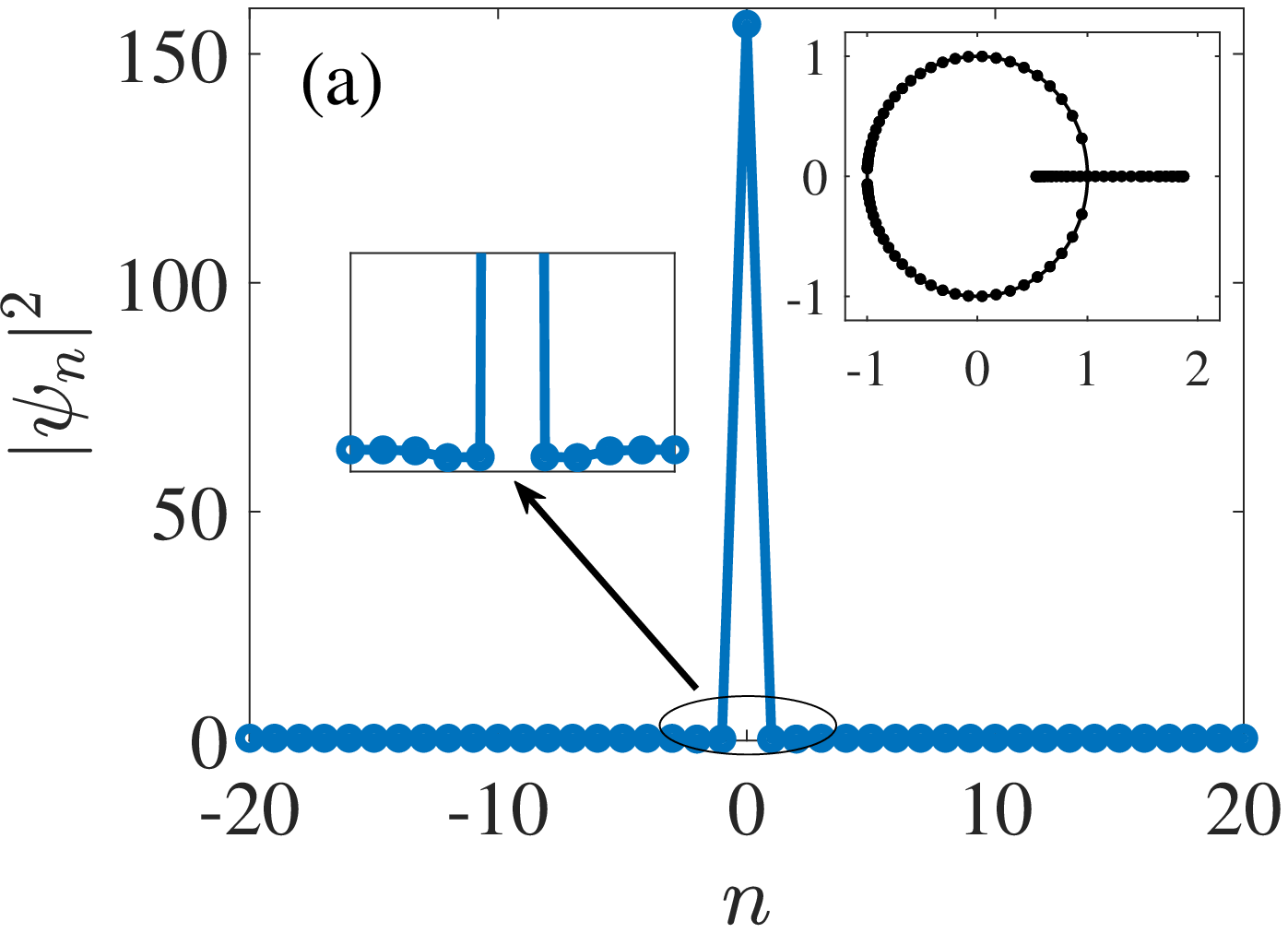}
\includegraphics[height=.18\textheight, angle =0]{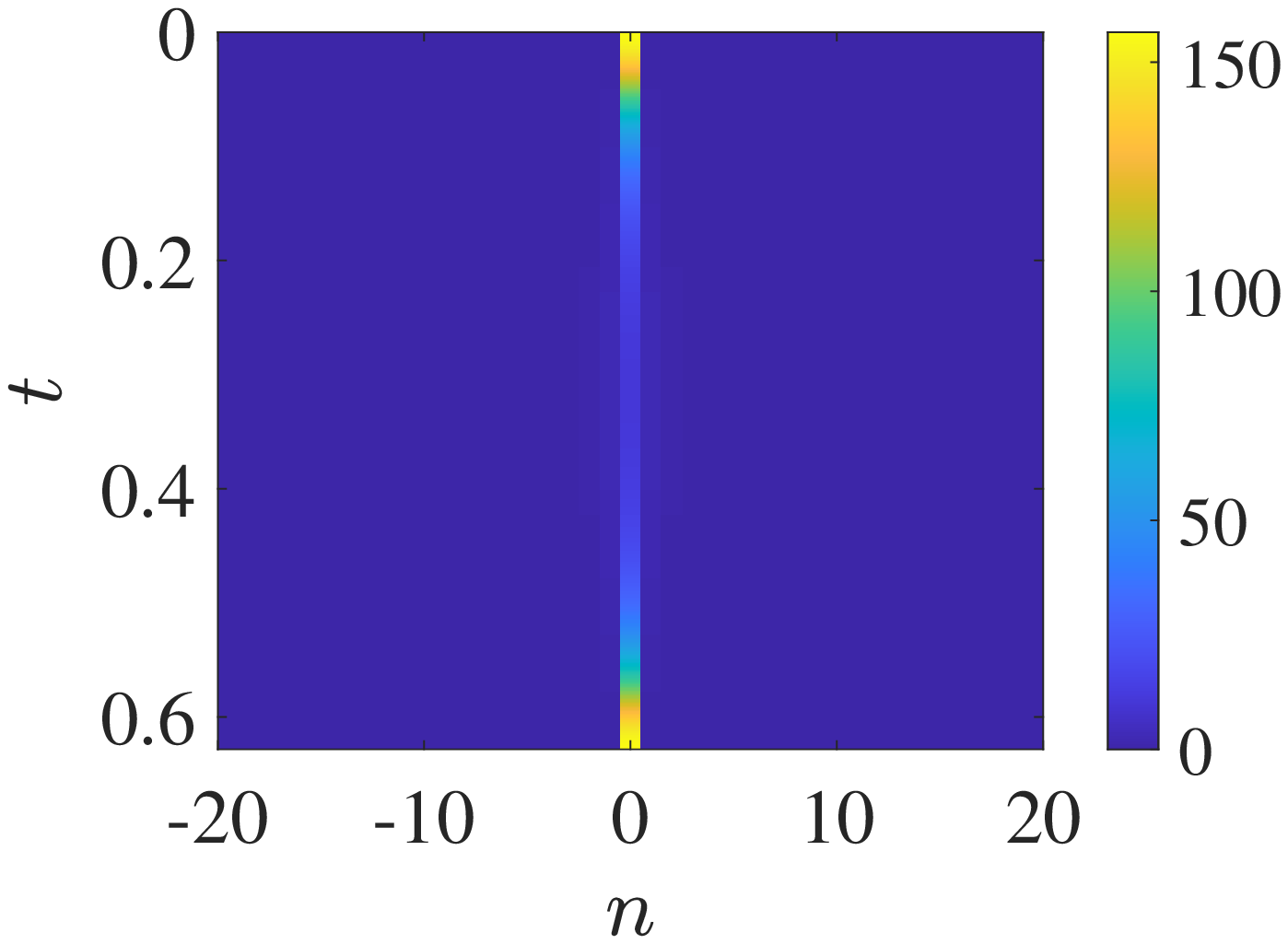}\\
\includegraphics[height=.18\textheight, angle =0]{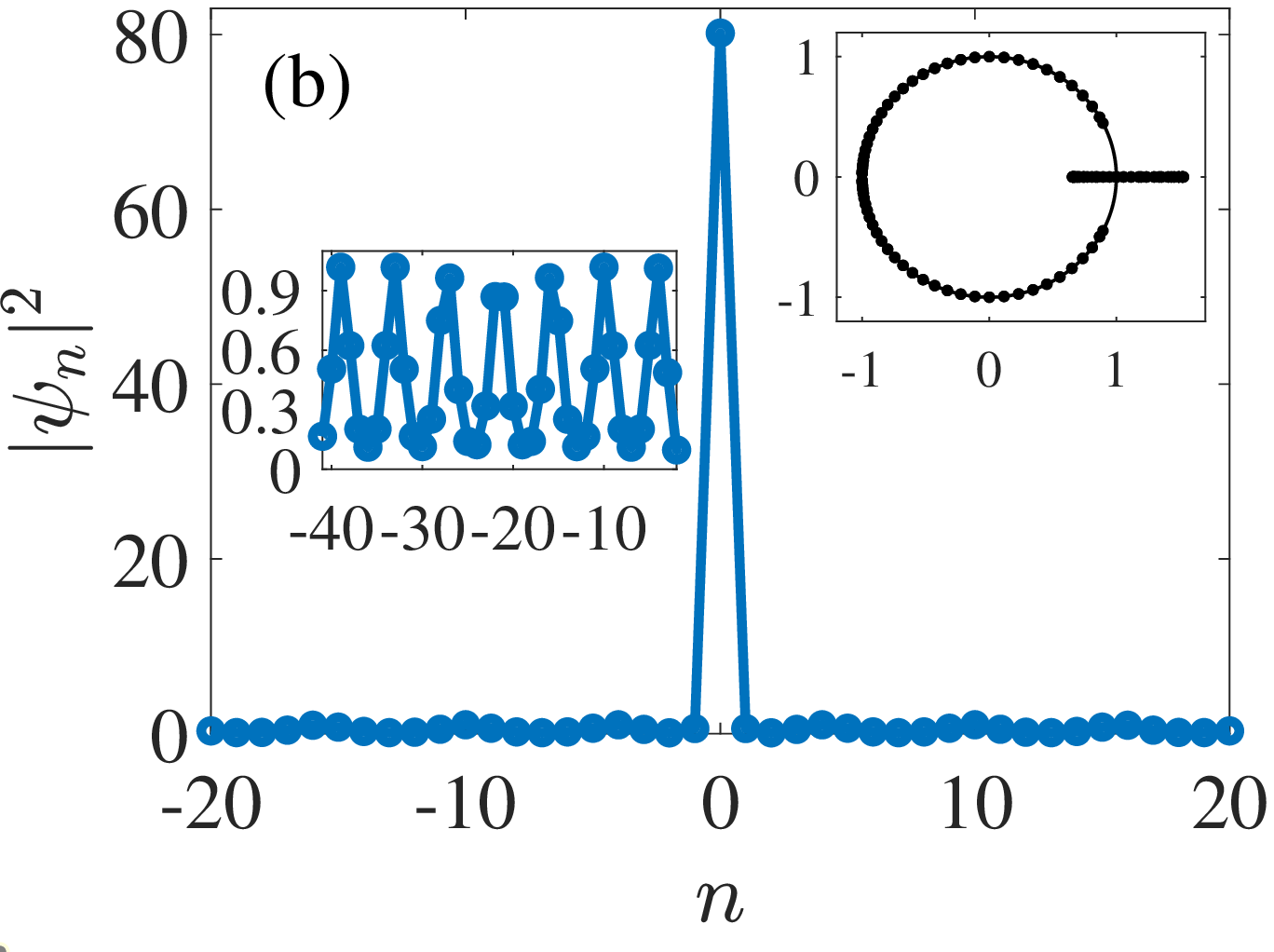}
\includegraphics[height=.18\textheight, angle =0]{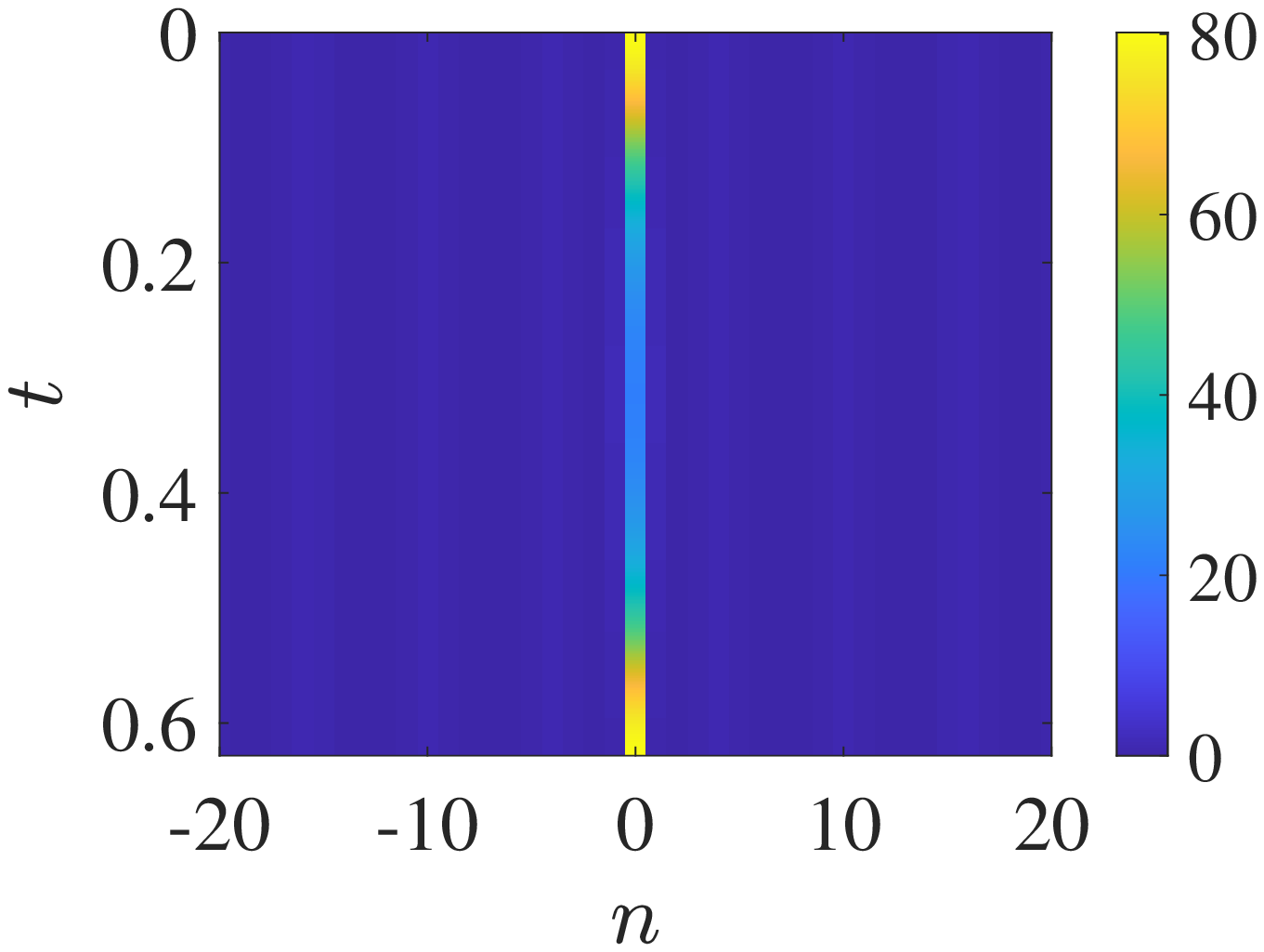}\\
\includegraphics[height=.18\textheight, angle =0]{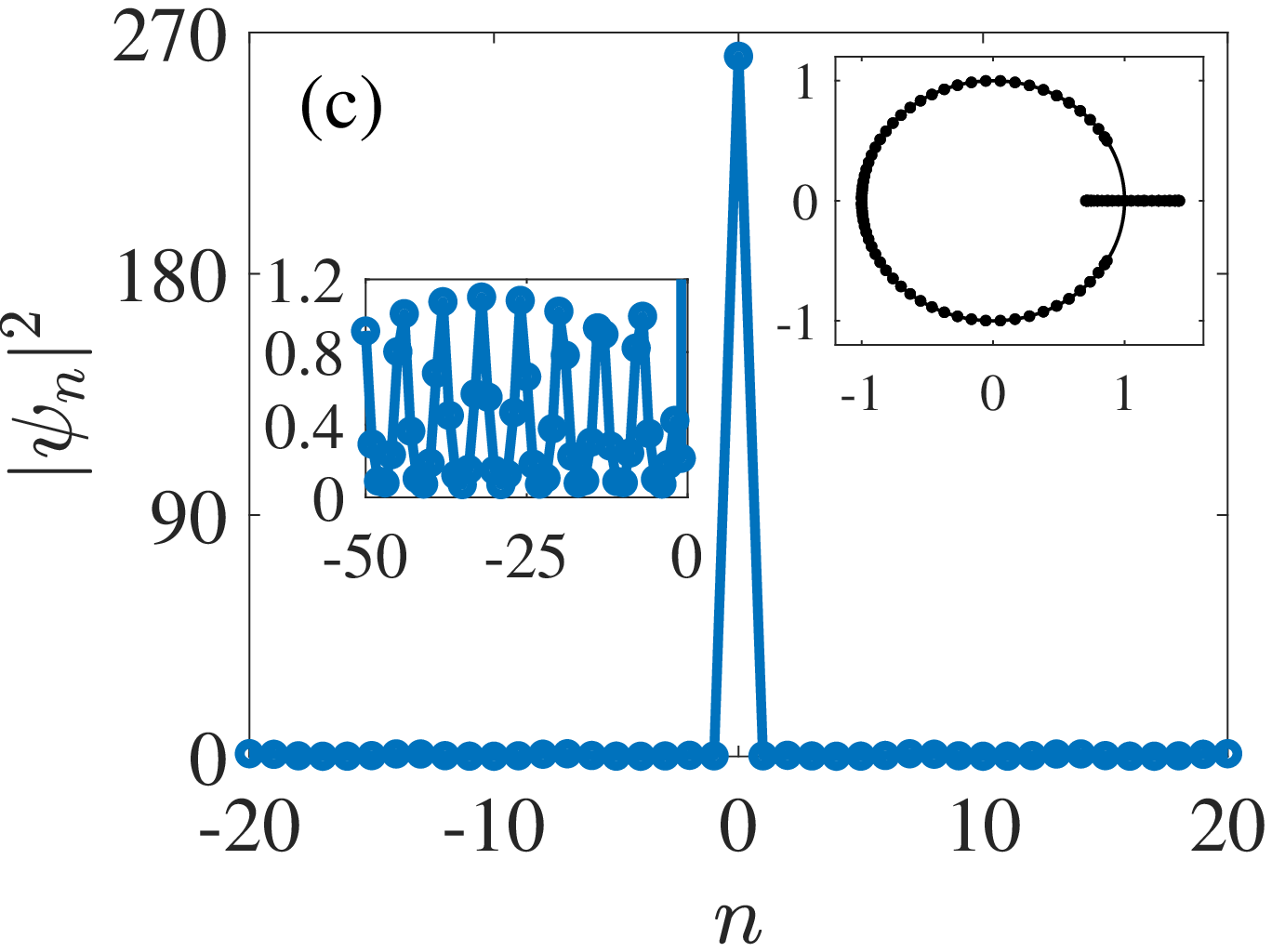}
\includegraphics[height=.18\textheight, angle =0]{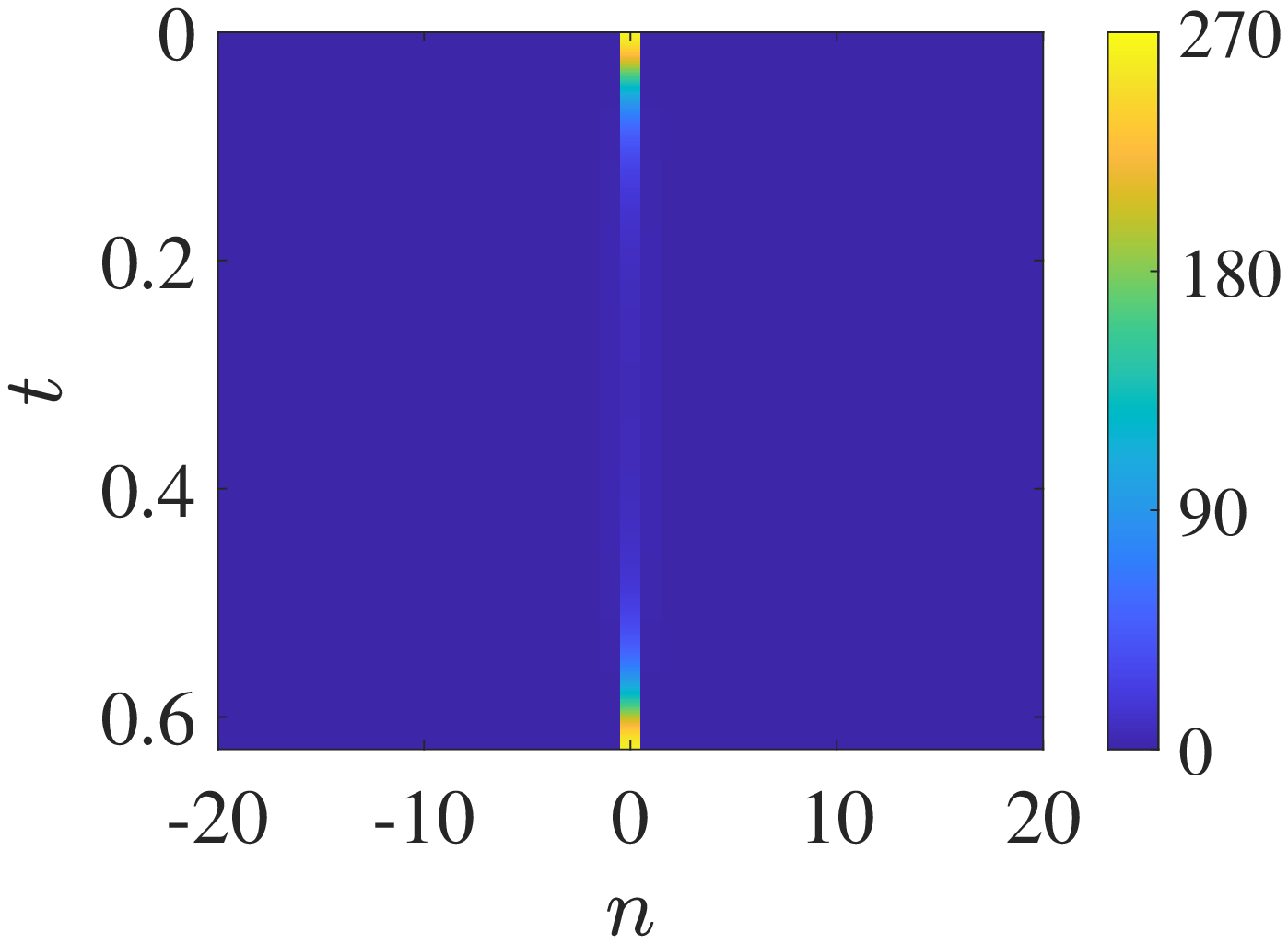}\\
\includegraphics[height=.18\textheight, angle =0]{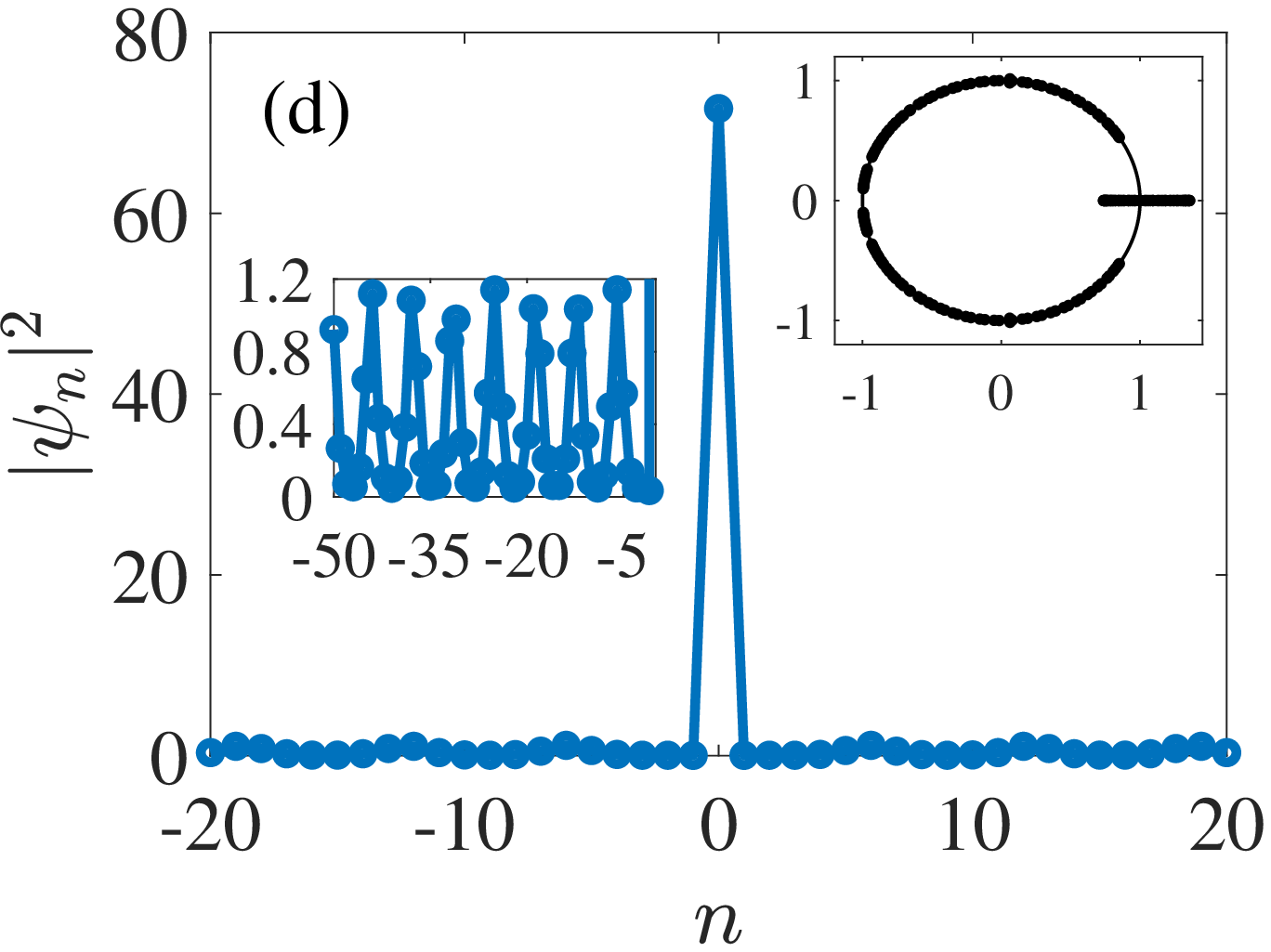}
\includegraphics[height=.18\textheight, angle =0]{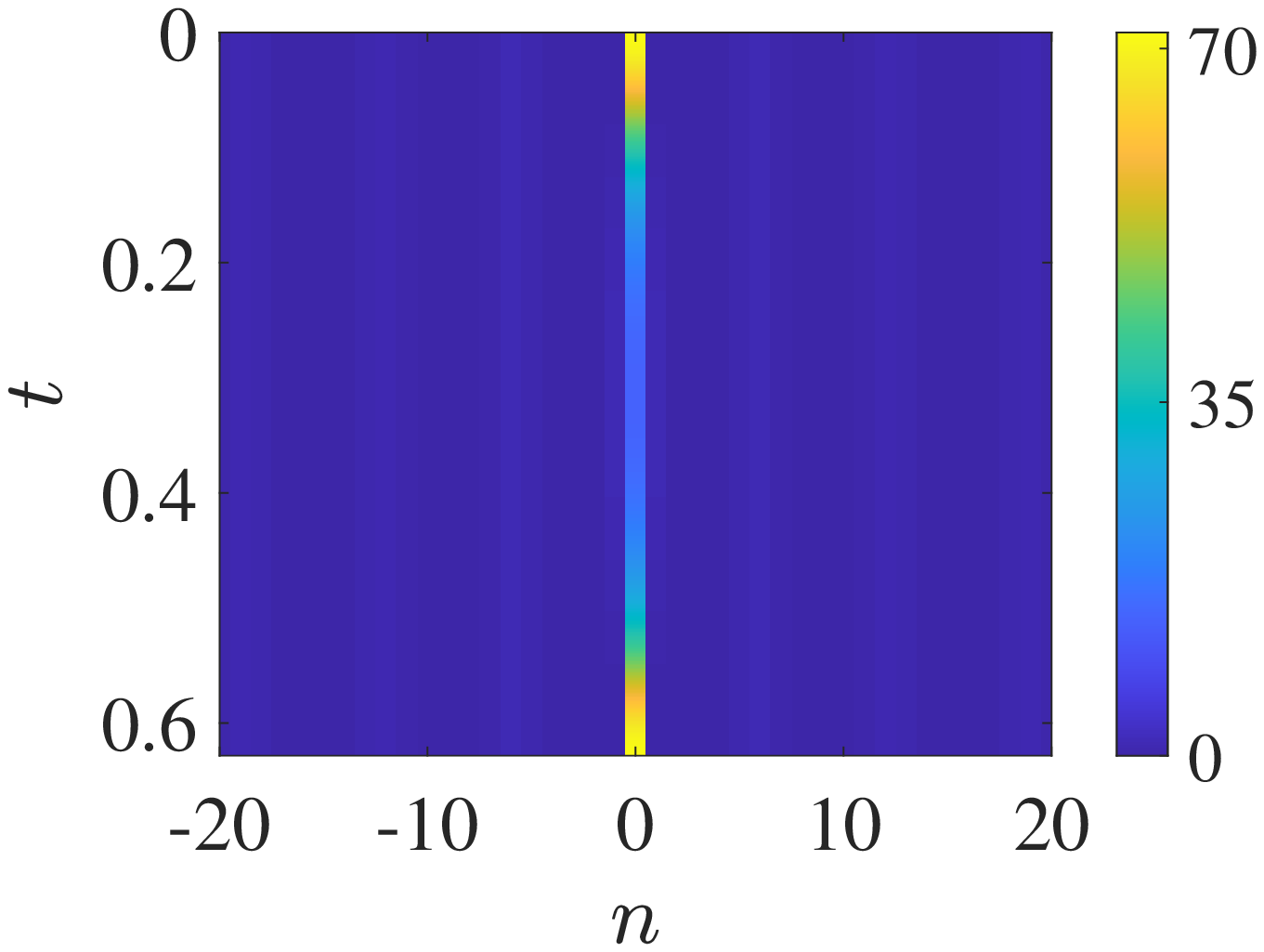}\\
\includegraphics[height=.18\textheight, angle =0]{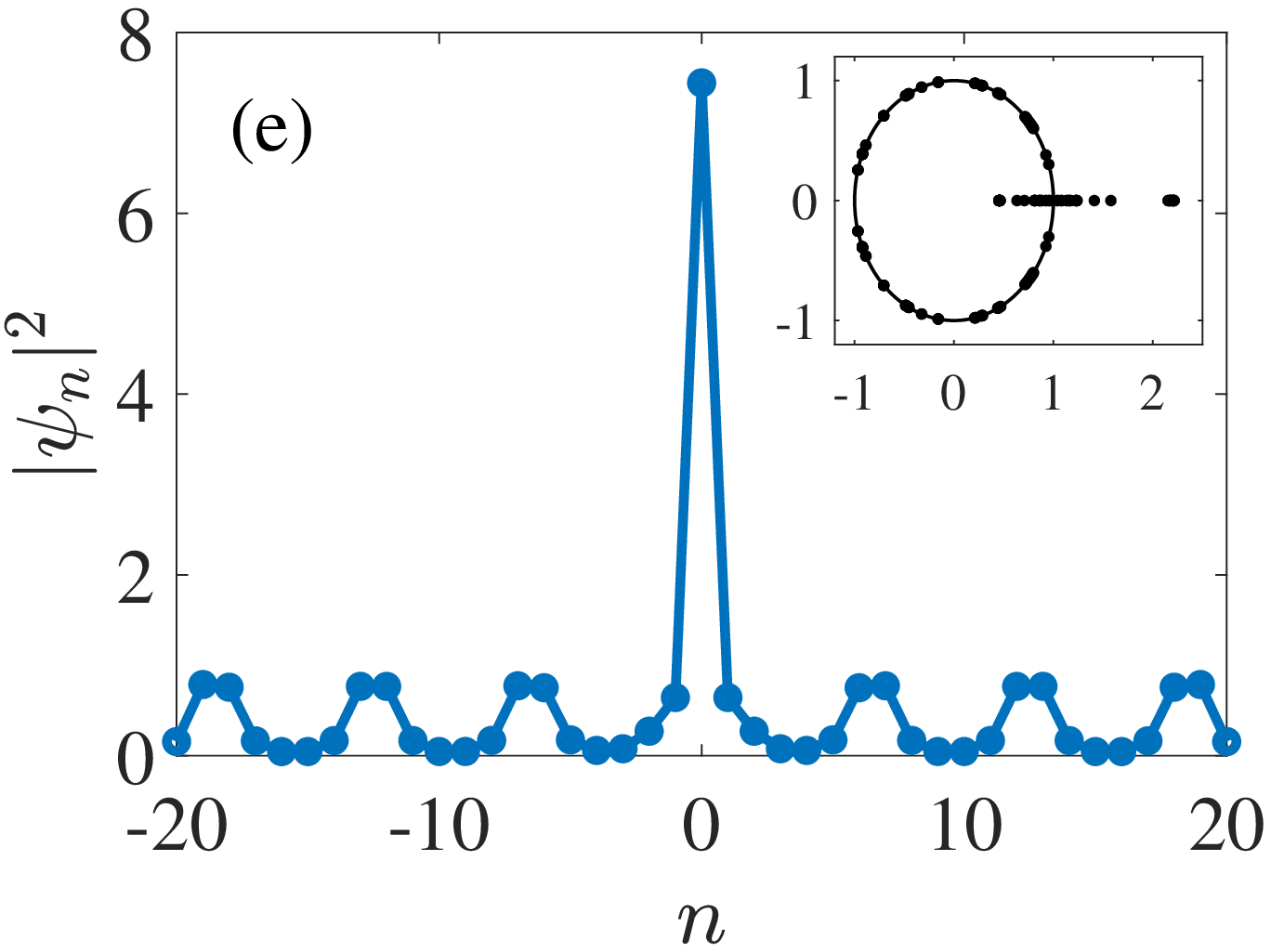}
\includegraphics[height=.18\textheight, angle =0]{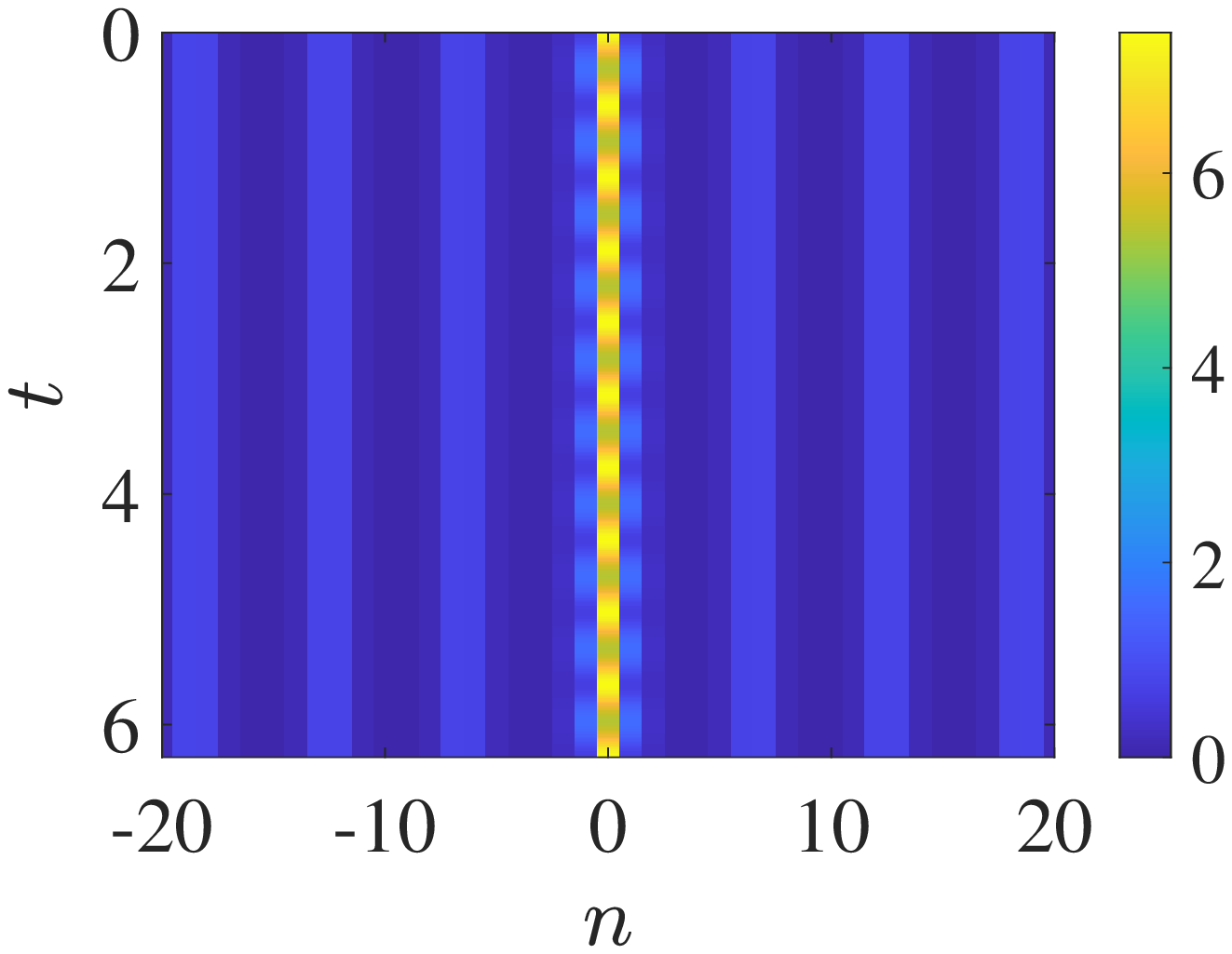}\\
\end{center}
\caption{(Color online) Same as Fig.~\ref{fig3} but for the case
of $\omega=10$ associated with the right panel of Fig.~\ref{fig2}. 
Again, the left and right columns correspond to the spatial distribution 
of $|\psi_{n}|^{2}$ and its spatio-temporal evolution, respectively. 
The integration was performed over one period for (a)-(d) and over 
$10$ periods for panel (e), respectively. The insets shown on the 
left and right of each panel provide close-ups of the profiles as 
well as Floquet spectra. Panels (a)-(c) correspond to $g=1$ whereas
the ones of (d) and (e) to $g=0.9$ and $g=0$ (DNLS case), respectively.
}
\label{fig4}
\end{figure}

Figure~\ref{fig2} summarizes our results on the existence of
of time-periodic solutions to the Salerno model [cf. Eq.~\eqref{salerno0}].
In particular, we demonstrate the dependence of $\textrm{max}\left(|\psi|\right)$
on $g$ for $\omega=5$, $\omega=25$, and $\omega=10$ in the left, 
middle, and right panels, respectively. Note that the labels (a)-(e) 
in the left and right panels are connected to Figs.~\ref{fig3} and~\ref{fig4},
respectively. Based on the panels of Fig.~\ref{fig2}, a cascade of 
turning points is clearly evident although all of them suggest an 
intriguing finding that we discuss now. We consider first the left panel 
of the figure corresponding to $\omega=5$. Starting from $g=1$, 
the time-periodic solution departs from the integrable limit (i.e., 
the AL limit), heading to smaller values of $g$ until it reaches a 
turning point at $g\approx 0.981$ upon which it \textit{comes back} 
to $g=1$, i.e., the AL limit, and then follows a ``snake'' pattern 
with \textit{several crossings} of the AL limit. It should be noted 
that we stopped our continuation algorithm at $g=1$ (the terminal 
solution is labeled by (e) therein). Based on our additional numerical 
investigations, this is en route to multiple additional crossings of 
$g=1$ (results are not shown). The labels of the left panel are connected 
with the prototypical configurations for this case in Fig.~\ref{fig3}. 
In particular, its left column presents the spatial distribution of the 
densities of the time-periodic solutions, i.e., $|\psi_{n}|^2$ for values 
of $g=1$ (panels (a), (c)-(e)), and $g=0.982$ (panel (b)), respectively. 
The insets therein correspond to the associated Floquet multipliers which 
themselves suggest that all computed solutions are highly unstable with a 
dominant unstable mode of the order of $\lambda_{r}\sim\mathcal{O}(1)$. 
Again, this can be naturally expected on the basis of the instability of 
the background. The striking feature of the solutions presented in panels 
(b)-(e) is the formation of an oscillatory background (of small amplitude) 
in contrast with the KM breather of panel (a) where the localized wave sits 
atop a constant background. Such profiles featuring small in-amplitude wave
trains are strongly reminiscent of \textit{nanoptera}, and to the best of 
our knowledge are first reported for the Salerno model in the present work 
(see, also the recent works of~\cite{mason_toda,peli_nano} for Toda lattices 
and the DNLS with saturation). The other striking feature of our findings is 
that the KM breather~\eqref{km_exact} of the AL model ($g=1$) seems \textit{not} 
to be the only solution at that limit, as this has already been evident in panels 
(b)-(e) of Fig.~\ref{fig3}. Although the investigation of those extra solutions 
by using integrable systems techniques is of fundamental importance, it is beyond 
the scope of our present work. We summarize our results in this case by presenting
the respective spatio-temporal dynamics in the right panel of Fig.~\ref{fig3}. In 
particular, contour plots of the spatio-temporal evolution of the density of time-periodic
solutions are shown for one period ($T=2\pi/5$ in this case). Though the solutions 
remain robust for one period, thus validating our numerical approach for identifying
them (via fixed-point iterations), we observed the emergence of the instability 
which happens at later times (results not shown) leading to a breakdown of the 
localized waveform.

Similarly, the middle panel of Fig.~\ref{fig2} corresponds to the case of $\omega=25$.
It presents a type of spiral structure around the KM solution that eventually leads 
to a progressively more dense loop structure that keeps repeating itself in the 
course of the pseudo-arclength continuation. See also the inset therein which suggests 
that complex structures may arise also at a fine scale within the bifurcation diagram.
The identified periodic states bear similar features as before, including progressively more 
pronounced (along the relevant branches) tails in the relevant waveforms. Furthermore, 
in line with earlier (MI-based and computational) findings and general expectations, the 
solutions are generically found to be unstable, due to the instability of their 
respective background.

It is perhaps important to highlight two pervading features regarding the nature 
of our results (both the ones above, as well as the ones that will follow). The 
first one is that to the best of our knowledge the results above (and below) 
constitute the first definitive identification of rogue-like patterns in discrete 
(nonlinear dynamical lattice) systems beyond the extremely important, yet practically 
limited realm of integrable systems. Secondly, our findings are not solely of interest 
to non-integrable dispersive system practitioners, but they are also a motivation for
further integrable system investigations. The panel (b) of Fig.~\ref{fig3}, for 
example, suggests the existence of KM-type solutions on top of a stationary nanopteronic 
background which is time-independent in the spirit of recent works of~\cite{peli1,peli2} 
in corresponding continuum limit problems. These are states that we believe are eminently 
relevant to explore in an analytical form within the framework of integrable systems 
(although this is outside the scope of the present study).

We conclude this section of our findings by going through the right panel of Fig.~\ref{fig2}
and the associated results shown in Fig.~\ref{fig4} corresponding to the case of $\omega=25$. 
Based on the former, we observe again a cascade of crossings happening at the AL limit (see, 
the inset therein), however here the bifurcation curve starts heading towards smaller values 
of $g$ until \textit{it reaches} $g=0$, i.e., the DNLS limit. Notice how this appears to be an
intermediate case between the less ordered and more expanded diagrams of small $\omega$ and the
more ordered and confined diagram of $\omega=25$. The profiles shown in panels (a)-(c) in 
Fig.~\ref{fig4} correspond to values of $g=1$ whereas the ones of (d) and (e) to $g=0.9$ and
$g=0$, respectively. Again, a common finding is that all time-periodic solutions identified in
this work feature an oscillatory background as soon as we depart from the integrable limit 
(see, the left insets in the left column of Fig.~\ref{fig4}) as well as they are unstable (see, 
the right insets therein). This effect is even more pronounced for the case with $g=0$ as is shown
in panel (e) of the figure. To the best of our knowledge, such time-periodic solutions of DNLS 
on a background have not appeared in the literature so far; indeed we are only aware of such 
effectively quasi-periodic solutions of the model on top of a vanishing background per the work
of~\cite{johaub}. This solution is also unstable as is evident in the Floquet spectrum shown in 
the inset of the panel, although it remains robust over many periods of time integration. Indeed, 
the right panel therein demonstrates the spatio-temporal evolution of the density again over $10$
periods where $|\psi_{0}(t)|^2$ breathes over time. It should also be noted that the oscillatory 
background is rather visible in this case (as well as the one of panels (b) and (d)) but again it
remains steady over the time evolution, i.e., corresponding to a definitive nanopteronic state
in this setting.

\section{Conclusions and Future Challenges}
\label{sec:conclusions}
In this work, we made an attempt to explore the existence, stability
and dynamics of time-periodic solutions of the Salerno model. This was 
with a three-fold scope in mind: firstly, to establish that relevant 
solutions such as the KM are not unique or particular to the integrable
limit, but can be continued to generic non-integrable values of the
homotopic parameter $g$. Secondly, we wished to explore whether additional 
intriguing solutions could arise in the integrable model, a feature 
that was brought forth from our pseudo-arclength continuation results 
for periodic orbits. Lastly, we intended to examine whether some of the 
relevant solutions could be continued to the DNLS limit of $g=0$; here, 
we found that for suitable choice of the breather frequency indeed that 
was also possible. Upon employing fixed-point methods, we identified the 
pertinent waveforms using Newton's method (for periodic orbits) and their 
stability was inferred by performing a Floquet analysis. The use of 
pseudo-arclength continuation allowed us to perform a parametric 
continuation over the homotopy parameter $g$ and this proved to be crucial 
in unraveling the complexity of the possible solutions in the Salerno model. 
Additionally for the integrable limit of $g=1$ we identified multiple time-periodic
solutions that sit atop of an oscillatory background being strongly reminiscent
of nanoptera. Another striking finding of our work was the time-periodic solution
which was identified at the DNLS limit, i.e., $g=0$. To the best of our knowledge,
such a waveform for the DNLS (on top of a non-vanishing background) was not previously
reported.

Based on the above findings and computational techniques that we have developed 
in this work, there are clearly many directions for future studies. At the AL
limit, a potential analysis of the Zhakarov-Shabat problem for the nanopteronic 
solutions reported in Section~\ref{sec:num_results} will be of paramount importance in order to 
derive them in possibly closed form (analogously to their identification in continuum 
problems~\cite{peli1,peli2}). On the other hand, it is worth investigating the $g=0$
solution reported in the present work, and in particular to study the configuration 
space of solutions as a function of $\omega$ in this case. This will pave the way towards
potentially identifying Peregrine-like entities for the DNLS as time-periodic solutions
at the limit $T\to\infty$ ($\omega\gg1$). Another important path to explore is the 
continuation of the present time-periodic solutions as the distance between adjacent
sites decreases, thus approaching the continuum limit when $C=1/h^{2}$ increases.
This way, we will be able to connect our findings with the results for the (focusing)
NLS. A subset of the above computational studies can be carried out quite efficiently
with the use of state-of-the-art bifurcation packages such as AUTO~\cite{auto} and
COCO~\cite{coco} which additionally allow branch switching, among other features.
Such directions are presently under consideration and will be reported in future
publications.

\begin{acknowledgments}
J.C.-M. was supported by MAT2016- 79866-R project (AEI/FEDER,UE). PGK
acknowledges support from the U.S.~National Science Foundation under
Grants no.~PHY-1602994 and DMS-1809074 (PGK). He also also acknowledges
support from the Leverhulme Trust via a Visiting Fellowship and the Mathematical
Institute of the University of Oxford for its hospitality during part of
this work.
\end{acknowledgments}

\end{document}